\documentclass[fleqn,usenatbib,useAMS]{mnras}

\usepackage{newtxtext,newtxmath}
\usepackage[T1]{fontenc}

\DeclareRobustCommand{\VAN}[3]{#2}
\let\VANthebibliography\thebibliography
\def\thebibliography{\DeclareRobustCommand{\VAN}[3]{##3}\VANthebibliography}

\usepackage{graphicx}
\usepackage{tabularx}
\usepackage{amsmath}	
\usepackage{booktabs}
\usepackage{mathtools,leftidx}

\title[Asteroseismic modelling of $\mu$ Herculis]{Glitch analysis and asteroseismic modelling of subgiant $\mu$ Herculis: confirming and interpreting the $\Gamma_1$ peak as the helium glitch}

\author[Gupta et al.]{Advik Gupta,$^{1}$\thanks{E-mail: advik.gupta.phy21@itbhu.ac.in (AG); kuldeepv89@gmail.com (KV); hans@phys.au.dk (HK)}
Kuldeep Verma,$^{1}$
Hans Kjeldsen,$^{2}$
Frank Grundahl,$^{2}$
J\o rgen Christensen-Dalsgaard,$^{2}$
\newauthor
Mark L. Winther,$^{2}$
Jakob L. R\o rsted,$^{2}$
Amalie Stokholm,$^{3}$
V\'{i}ctor Aguirre B\o rsen-Koch,$^{4}$
Pere L. Pall\'{e}$^{5,6}$\\
$^{1}$Department of Physics, Indian Institute of Technology (BHU), Varanasi 221005, India\\
$^{2}$Stellar Astrophysics Centre (SAC), Department of Physics and Astronomy, Aarhus University, Ny Munkegade 120, DK-8000 Aarhus, Denmark\\
$^{3}$School of Physics and Astronomy, University of Birmingham, Edgbaston, Birmingham B15 2TT, UK\\
$^{4}$DARK, Niels Bohr Institute, University of Copenhagen, Jagtvej 128, 18, 2200, Copenhagen, Denmark\\
$^5$Instituto de Astrof\'{i}sica de Canarias, 38200 La Laguna, Tenerife, Spain\\
$^6$Universidad de La Laguna (ULL), Departamento de Astrof\'{i}sica, 38206 La Laguna, Tenerife, Spain
}

\date{Accepted XXX. Received YYY; in original form ZZZ}

\pubyear{\the\year{}}

\begin{document}
\label{firstpage}
\pagerange{\pageref{firstpage}--\pageref{lastpage}}
\maketitle


\begin{abstract}
The measurements of pressure-mode and mixed-mode oscillation frequencies in subgiant stars offer a unique opportunity to probe their internal structures -- from the surface to the deep interior -- and to precisely determine their global properties. We have conducted a detailed asteroseismic analysis of the benchmark subgiant $\mu$ Herculis using eight seasons of radial velocity observations from the SONG-Tenerife, and have determined its mass, radius, age, and surface helium abundance to be $1.105_{-0.024}^{+0.058}$ M$_\odot$, $1.709_{-0.015}^{+0.030}$ R$_\odot$, $8.4_{-0.1}^{+0.4}$ Gyr, and $0.242^{+0.006}_{-0.021}$, respectively. We have demonstrated that simultaneously fitting the helium glitch properties, oscillation frequencies, and spectroscopic observables yields a more accurate inference of the surface helium abundance and hence stellar age. A significant discrepancy between the observed extent of the helium ionization zone and that predicted by stellar models is identified and examined, underscoring potential limitations in the current modelling of stellar interiors. Our analysis confirms that the helium glitch originates from the region between the two stages of helium ionisation, i.e. from the $\Gamma_1$ peak, rather than from the second helium ionisation zone itself. Within the conventional formalism, this implies that the glitch analysis characterises the region located between the two helium ionisation zones. 

\end{abstract}





\section{Introduction}
  
The measurement of pressure-mode (or p-mode) frequencies in solar-like main-sequence stars has become a routine and powerful tool for probing their internal structure and dynamics, as well as for determining their fundamental properties. This progress has been made possible primarily due to high-precision photometric data from space-based missions such as CoRoT \citep{bagl09}, {\it Kepler/K2} \citep{boru09,howe14}, and TESS \citep{rick14}. In contrast, gravity modes (or g-modes) have not yet been detected in solar-type main-sequence stars, as their g-mode cavities are located deep within the stellar interior. Despite numerous efforts, robust detections of g-modes remain elusive even in the Sun \citep[see][and references therein]{schu18}, thereby limiting our ability to investigate the solar and stellar cores in detail. However, once stars evolve off the main sequence and enter the subgiant phase, their p-mode and g-mode cavities are sufficiently close for these two fundamentally different types of modes to couple, resulting in so-called mixed modes. These mixed modes have been observed and now serve as crucial diagnostics for probing the deep interiors of evolved stars.

Mixed-mode frequencies are highly sensitive to a star’s internal structure and evolutionary state \citep[see e.g.,][]{stok19,li19,chap20,clar25}. Their detection in subgiants offers a powerful means of tightly constraining the physical conditions in their interior and evolutionary state. In principle, comparing the observed p-mode and mixed-mode frequencies with those predicted by theoretical stellar models enables rigorous testing and refinement of stellar evolution models, as well as precise inference of fundamental stellar parameters. However, in practice, modelling such stars poses significant challenges, as it requires either on-the-fly modelling \citep[see e.g.,][]{farn25} or a densely populated grid of evolutionary tracks with finely sampled models along each track. Although sophisticated genetic algorithm–based methods of the former kind are well-suited to this problem, they are generally computationally expensive, as the models computed for a particular star cannot be reused for others. On the other hand, the latter grid-based modelling approaches typically suffer from inadequate grid resolution. Hence, interpolation across the grid is frequently adopted, though its reliability is constrained by the reported discrepancies exceeding 1~$\mu$Hz between the interpolated and computed mixed-mode frequencies \citep{li19,clar25}.

It is well known that the physical description of the near-surface layers of stars has shortcomings. As a result, one-dimensional stellar evolution models systematically predict oscillation frequencies that deviate from observations in a frequency-dependent manner -- an effect commonly referred to as the surface effect \citep[see e.g.,][]{chris91}. To mitigate this issue in stellar modelling, several empirical corrections to the model frequencies have been proposed \citep{hans08,ball14,sono15}. However, significant discrepancies often persist between observed and model frequencies, leading to large chi-square values and unrealistically low initial helium abundances inferred \citep[see e.g.,][]{math12}. 

To improve the accuracy of theoretical models, \citet{jorg18} introduced an approach in which the outermost layers of one-dimensional stellar models are replaced by horizontally averaged structures derived from more realistic three-dimensional hydrodynamic simulations \citep[see also][]{mosu20,jorg21,zhou25}. Although this method improves the agreement between observed and model frequencies, notable differences still remain. 

An alternative strategy involves using frequency separation ratios in the modelling process, as these are less sensitive to the near-surface layers \citep[see e.g.,][]{roxb03,oti05}. However, while this approach reduces the sensitivity to the surface effect, it offers less constraining power compared to the direct use of the oscillation frequencies. Moreover, this technique is not applicable to subgiants whose oscillation spectrum contains mixed modes.

The region of partial ionisation of helium in solar-like stars leaves a subtle yet detectable imprint on their oscillation frequencies \citep[see e.g.,][]{mazu14,verma17}. In particular, it leaves a characteristic oscillatory signal in the mode frequencies as a function of the radial order \citep[see e.g.,][]{gough90,mont98,houd07}. Traditionally, this signature has been attributed to a localised depression or a glitch in the first adiabatic index, $\Gamma_1$, caused by the second helium ionisation zone. However, in their investigation of the potential for asteroseismic inference of the helium abundance in red-giant stars, \citet{broom14} analysed theoretical oscillation frequencies and unexpectedly found that the helium-induced signature originates from a region located between the two helium ionisation zones, where $\Gamma_1$ exhibits a local maximum. A similar conclusion was reached by \citet{verma14b} through a detailed theoretical study of main-sequence stars. Recently, a new formalism based on an analytic description of the $\Gamma_1$ profile for analysing the helium glitch signature was developed by \citep{houd21,houd22}.

In this study, we model the asteroseismic benchmark G5 subgiant $\mu$ Herculis, observed by the Hertzsprung Stellar Observations Network Group (SONG) node at the Observatorio del Teide on Tenerife, Spain, during the period 2014-2021 (a total of eight seasons), to infer its fundamental stellar properties. Our analysis uses individual oscillation frequencies along with the observed helium glitch properties. Since the initial helium abundance, $Y_i$, is anticorrelated with mass, $M$ \citep[see e.g.,][]{metc09,lebr14,verma16}, the constraint on the surface helium abundance, $Y_s$, from the helium glitch observables allows for precise and accurate estimates of the star’s mass and age. We compare our results with those reported by \citet[][hereafter G17]{grun17}, who analysed data from an earlier, shorter observational campaign spanning 215 nights. In addition, we conduct a detailed investigation into the physical origin of the helium glitch signature in this subgiant.

The paper is organised in the following order. We briefly describe the observational data we have for $\mu$ Herculis in Section~\ref{sec:observed_data}. The details of the stellar model grids and the modelling approach are provided in Section~\ref{sec:stellar_modeling}. In Section~\ref{sec:results}, we present our results. The conclusions of the paper are summarised in Section~\ref{sec:conclusions}.

\begin{table} 

    \caption{Various observed parameters for our target, $\mu$ Herculis.}
    \label{tab:table_observed}
    
    \begin{tabularx}{\columnwidth}{X c X}
    
        \hline
        \hline
        Parameter & Value & Reference \\[0.5ex]
        \hline 
        \noalign{\vskip 0.1cm}

        [Fe/H] [dex] & $0.28 \pm 0.07$ & \citet{grun17} \\
        $T_{\rm eff}$ [K] & $5560 \pm 80$ & \citet{grun17} \\

        \noalign{\vskip 0.1cm}
        \hline
        \noalign{\vskip 0.1cm}

        $R [R_{\odot}]$ & $1.73 \pm 0.02$ & \citet{grun17} \\
        $L \ [L_{\odot}]$ & $2.54 \pm 0.08$ & \citet{grun17} \\
        
        \noalign{\vskip 0.1cm}
        \hline 
        \noalign{\vskip 0.1cm}
        
        $\Delta\nu$ [$\mu$Hz] & $64.18 \pm 0.27$ & This work \\
        $\nu_{\max}$ [$\mu$Hz] & $1216 \pm 3$ & This work \\
        
        \noalign{\vskip 0.1cm}
        \hline 
        \noalign{\vskip 0.1cm}
        
        $\langle A_{\rm He} \rangle$ [$\mu$Hz] & $0.46 \pm 0.03$  & This work\\
        $\Delta_{\rm He}$ [s]& $92 \pm 18$ & This work \\
        $\tau_{\rm He}$ [s] & $1480 \pm 44$ & This work \\
        $\phi_{\rm He}$ & $3.99\pm0.59$ & This work \\
        
        \noalign{\vskip 0.1cm}
        \hline
        
    \end{tabularx}
\end{table}

\begin{table}
    \centering
    \caption{The individual oscillation frequencies of $\mu$ Herculis, along with their associated uncertainties, as measured from the SONG-Tenerife data.}

    \label{tab: table_freq}
        \begin{tabularx}{\columnwidth}{>{\centering\arraybackslash}X 
                                >{\centering\arraybackslash}X 
                                >{\centering\arraybackslash}X 
                                >{\centering\arraybackslash}X}
        \hline
        \raisebox{-1ex}{$\ell$} & \raisebox{-1ex}{$n$} & $\nu_{nl}$ & $\sigma_{nl}$ \\
          &   & ($\mu$Hz) & ($\mu$Hz) \\
        \hline
        $0^{*}$ & 8 &  608.124 & 0.122 \\
        0 & 9 &  672.315 & 0.145 \\
        0 & 10&  737.197 & 0.131 \\
        0 & 11&  800.832 & 0.105 \\
        0 & 12&  864.823 & 0.131 \\
        0 & 13&  928.629 & 0.059 \\
        0 & 14&  991.178 & 0.102 \\
        0 & 15& 1054.932 & 0.050 \\
        0 & 16& 1118.884 & 0.075 \\
        0 & 17& 1183.528 & 0.088 \\
        0 & 18& 1247.894 & 0.040 \\
        0 & 19& 1311.803 & 0.117 \\
        0 & 20& 1376.604 & 0.152 \\
        0 & 21& 1441.586 & 0.126 \\
        0 & 22& 1507.494 & 0.236 \\
        0 & 23& 1573.050 & 0.545 \\
        1 & 7 &  569.785 & 0.169 \\
        1 & 8 &  631.133 & 0.145 \\
        1 & 9 &  706.480 & 0.149 \\
        1 & 10&  766.228 & 0.122 \\
        1 & 11&  824.458 & 0.129 \\
        1 & 12&  903.838 & 0.118 \\
        1 & 13&  958.808 & 0.086 \\
        1 & 14& 1020.382 & 0.071 \\
        1 & 15& 1083.671 & 0.063 \\
        1 & 16& 1147.361 & 0.052 \\
        1 & 17& 1211.222 & 0.063 \\
        1 & 18& 1275.059 & 0.053 \\
        1 & 19& 1339.662 & 0.089 \\
        1 & 20& 1404.184 & 0.125 \\
        1 & 21& 1467.641 & 0.240 \\
        1 & 22& 1533.199 & 0.135 \\
        1 & 23& 1600.701 & 0.188 \\
        $1^{*}$ & 24& 1668.709 & 0.134 \\
        $1^{*}$ & 25& 1734.986 & 0.116 \\
        2 & 7 &  601.472 & 0.169 \\
        2 & 8 &  665.635 & 0.164 \\
        2 & 9 &  731.235 & 0.162 \\
        2 & 10&  795.731 & 0.196 \\
        2 & 11&  858.656 & 0.075 \\
        2 & 12&  923.138 & 0.090 \\
        2 & 13&  986.359 & 0.193 \\
        2 & 14& 1049.380 & 0.092 \\
        2 & 15& 1113.269 & 0.126 \\
        2 & 16& 1178.843 & 0.077 \\
        2 & 17& 1242.707 & 0.098 \\
        2 & 18& 1307.074 & 0.108 \\
        2 & 19& 1371.979 & 0.166 \\
        2 & 20& 1436.985 & 0.212 \\
        2 & 21& 1503.169 & 0.243 \\
        2 & 22& 1570.134 & 0.257 \\
        $3^{*}$ & 10 &  815.704 & 0.285 \\
        $3^{*}$ & 14 & 1075.203 & 0.095 \\
        $3^{*}$ & 15 & 1139.515 & 0.137 \\
        $3^{*}$ & 16 & 1203.920 & 0.110 \\
        $3^{*}$ & 17 & 1268.465 & 0.065 \\
        $3^{*}$ & 18 & 1333.213 & 0.132 \\
        $3^{*}$ & 19 & 1399.365 & 0.245 \\
        $3^{*}$ & 20 & 1464.445 & 0.316 \\
        \cline{1-4}

    \end{tabularx}
    
    \begin{flushleft}
    \footnotesize{$^{*}$ The modes excluded in the modelling (see text for details).}
    \end{flushleft}

\end{table}

\begin{figure}
    \centering
    \includegraphics[width=1.0\linewidth]{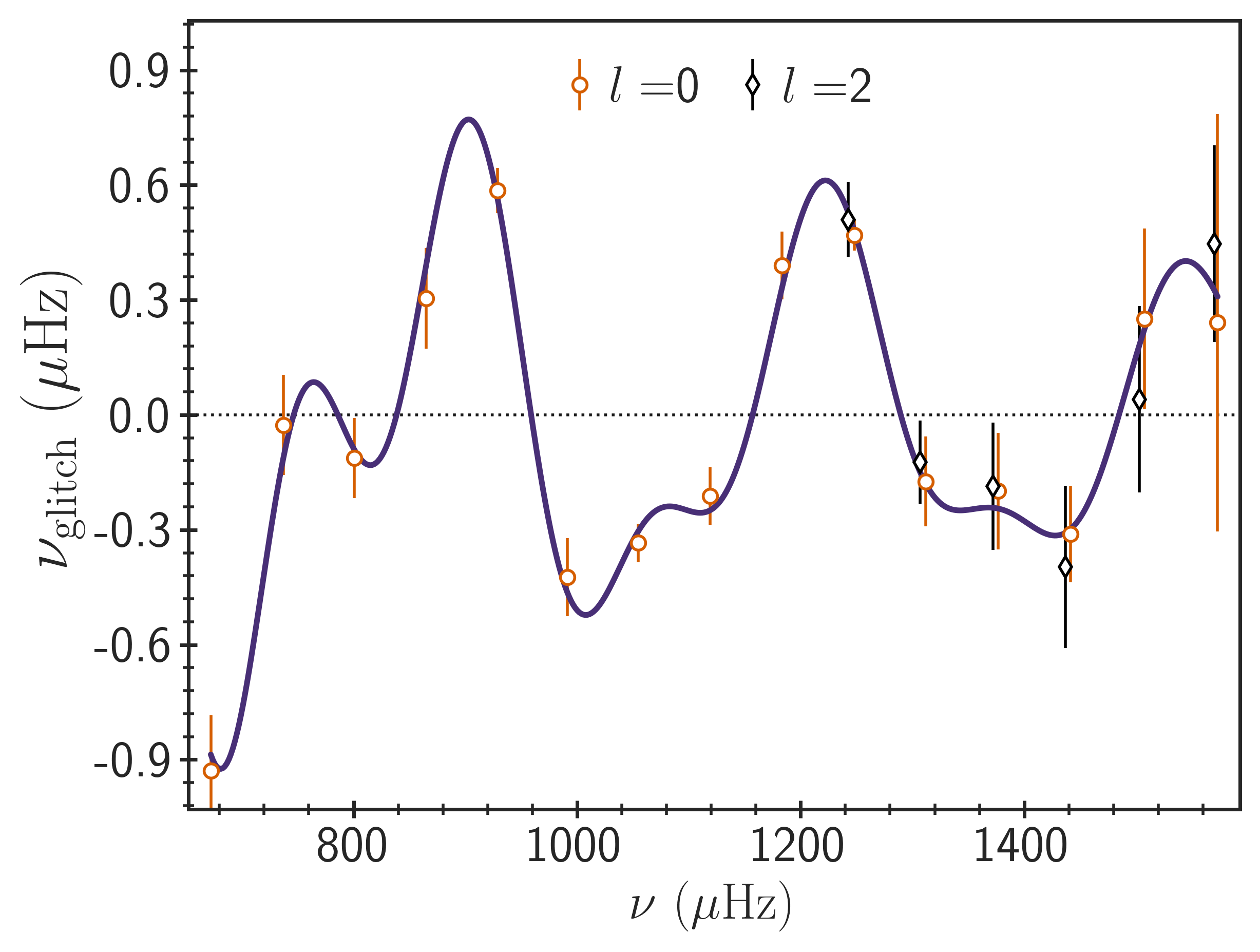}
        \caption{Acoustic glitch signatures (sum of both the helium and the base of convective envelope) as a function of frequency for $\mu$ Herculis. The circles and diamonds with errorbar represent the observed radial and quadrupole mode frequencies, respectively. The solid curve is the best fit to the data. The dotted horizontal line marks the zero level.}
    \label{fig:glitch fit}
\end{figure}


\section{Observational data}
\label{sec:observed_data}

We used the effective temperature, $T_{\rm eff}$, and the surface metallicity, $[{\rm Fe}/{\rm H}]$, for $\mu$ Herculis from G17, as listed in Table~\ref{tab:table_observed}. G17 adopted these values from \citet{jofre15}. Since the measured parameters in the literature differ significantly, they used inflated uncertainties as suggested by \citet{brun10}. 

Although the measured luminosity, $L$, and radius, $R$, of $\mu$ Herculis will not be used as constraints in stellar modelling, our best-fitting models will be evaluated against them for consistency, and hence we list them in Table~\ref{tab:table_observed}. Assuming no extinction, G17 estimated the luminosity using the measured visual magnitude \citep[][]{bess00} and the {\it Hipparcos} parallax. The radius is based on the angular diameter measurements using the Precision Astronomical Visual Observations beam combiner \citep{irel08}. The angular diameter was determined by them using a linear limb darkening law.

\begin{figure*}
    \centering
    \includegraphics[width=0.8\linewidth]{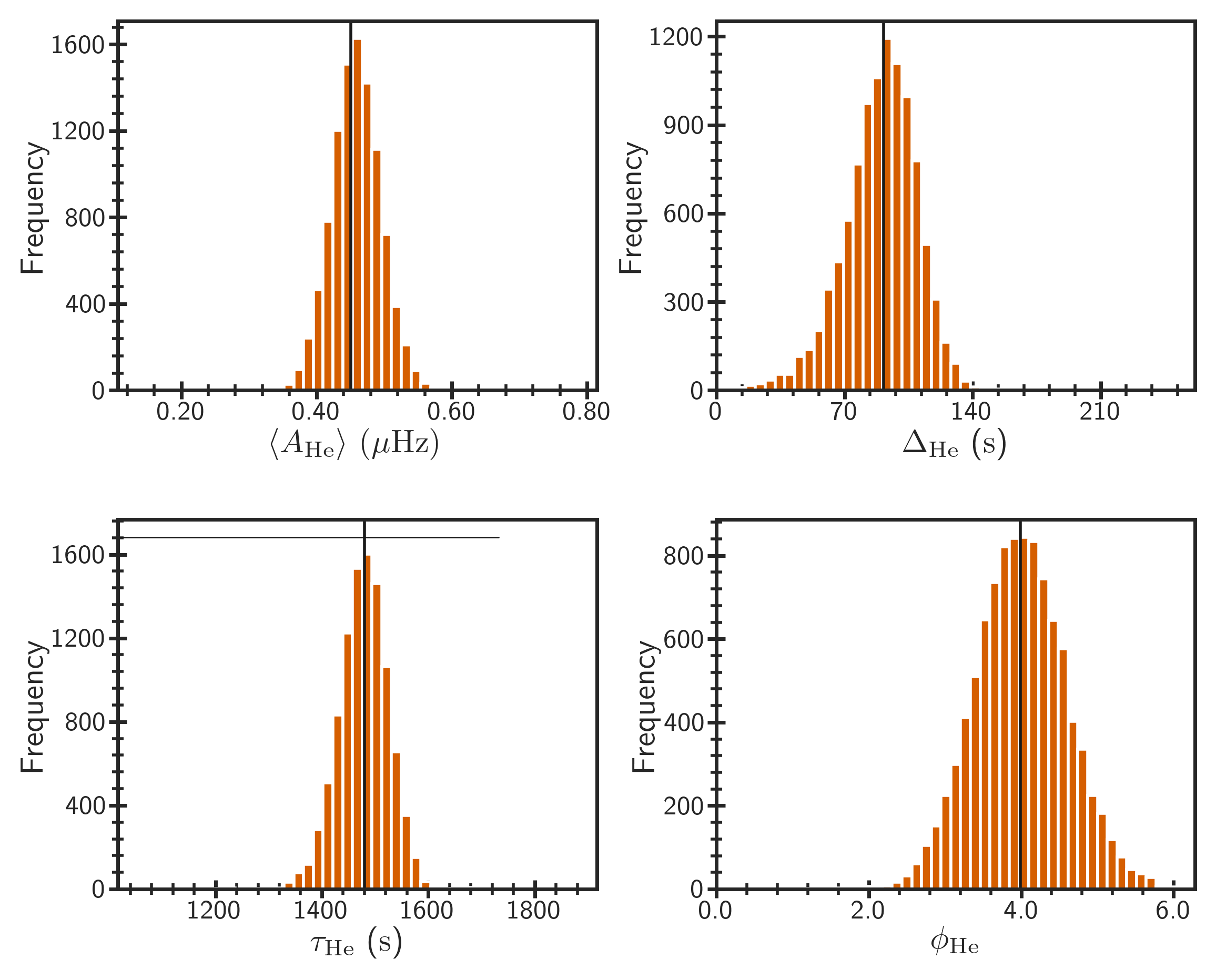}
    \caption{The distributions of helium glitch parameters obtained from Monte Carlo simulations with 10,000 realisations of the observed frequencies of $\mu$ Herculis. The top left, top right, bottom left, and bottom right panels show distributions for the average amplitude, acoustic width of the helium ionisation zone, its acoustic depth, and the phase, respectively. The vertical black line in each panel shows the median of the distribution. The horizontal line in the bottom left panel shows the range of acoustic depth explored for finding the global minimum.}
    \label{fig:glitch_realisations}
\end{figure*}

SONG-Tenerife has observed $\mu$ Herculis since 2014. A detailed analysis of its radial velocity time series acquired during 2014-2021 was performed. The time series was first processed by removing bad points, estimating point-by-point uncertainties, and applying a high-pass filter to suppress long-term trends. The power spectrum was then computed using the iterative sine-wave fitting (also known as prewhitening), with the uncertainties as the weights. The oscillation frequencies were extracted from the power spectrum by measuring the strongest peaks. This procedure is consistent with the approach adopted by G17. The details of the technique used to calculate the oscillation frequencies from the time series data are discussed in \citet{hans25}. The measured frequencies of $\mu$ Herculis are listed in Table~\ref{tab: table_freq}. Although several octupole modes are observed, we shall use only radial, dipole, and quadrupole modes in the present study. The measured values of the frequency of maximum power, $\nu_{\rm max}$, and the large frequency separation, $\Delta\nu$, are given in Table~\ref{tab:table_observed}.

The signal-to-noise (S/N) in the power spectrum decreases as we move away from $\nu_{\rm max}$, resulting in a typically monotonic increase in the errors on the frequencies. In other words, the frequencies at the two extreme ends are relatively more unreliable. We note that the lowest radial-mode frequency and the two highest dipole-mode frequencies have smaller errors compared to the corresponding neighboring frequencies (see Table~\ref{tab: table_freq}), and hence they are likely to have underestimated uncertainties. Since the helium glitch analysis presented below is highly sensitive to the lowest-frequency modes -- where the amplitude of the helium signature increases exponentially -- even small systematic uncertainties in these frequencies can significantly bias the inferred helium glitch parameters. To avoid such biases, we excluded these three modes from the analysis. 


\subsection{Inference of helium glitch properties}

The measured oscillation frequencies contain rich information about the stellar interior, including the region associated with the helium ionisation. It is believed that the localised dip in the profile of the first adiabatic index in the second helium ionisation zone causes a glitch in the acoustic structure of the star. \citet{houd07} assumed a Gaussian profile for the relative dip in $\Gamma_1$ \citep[see also,][]{gough02}:
\begin{equation}
   \label{eq:1}
    \frac{\delta\Gamma_1}{\Gamma_1} = -\frac{1}{\sqrt{2\pi}}\frac{\Gamma_{\rm II}}{\Delta_{\rm He}} e^{-(\tau-\tau_{\rm He})^2/2\Delta_{\rm He}^{2}},
\end{equation}
where $\Gamma_{\rm II}$ characterises the area of the dip, while $\Delta_{\rm He}$ and $\tau_{\rm He}$ denote its acoustic width and acoustic depth, respectively. Using a variational principle \citep[][]{chan63,lynd67}, they derived the contribution of the helium glitch to the oscillation frequency as: 
\begin{equation}
    \label{eq:2}
    \delta\nu_{\rm He} = A_{\rm He}\nu e^{-8\pi^{2}\Delta^{2}_{\rm He}\nu^{2}}\rm sin(4\pi\tau_{\rm He}\nu\ + \phi_{\rm He}),
\end{equation}
where $A_{\rm He}$ is proportional to $\Gamma_{\rm II}$, and $\phi_{\rm He}$ is a phase constant. Since the amplitude averaged over the observed frequency range, between $\nu_1$ and $\nu_2$, is defined as,

\begin{equation}
    \langle A_{\rm He
    }\rangle = \frac{\int^{\nu_2}_{\nu_1} A_{\rm He}\nu e^{-8\pi^2\Delta_{\rm He}^2\nu^2}d\nu}{\int^{\nu_2}_{\nu_1}d\nu}
\end{equation}

\begin{equation}
   \hspace{0.73cm} = \frac{A_{\rm He}[e^{-8\pi^2\Delta_{\rm He}^2\nu^2_1} - e^{-8\pi^2\Delta_{\rm He}^2\nu^2_2}]}{16\pi^2\Delta_{\rm He}^2[\nu_2-\nu_1]},
\end{equation}
it correlates better with the surface helium abundance \citep[][]{basu04,verma14a,verma19}, we use $\langle A_{\rm He} \rangle$, $\Delta_{\rm He}$ and $\tau_{\rm He}$ as our helium glitch observables. Note that although the observed $\phi_{\rm He}$ contains useful information about the stellar structure, we do not use it in the modelling because the model phase can be significantly influenced by the surface effect.

These helium glitch observables serve as an additional set of constraints in the stellar model fitting, allowing robust determination of the surface helium abundance. This breaks the degeneracy between $Y_s$ and other stellar parameters such as mass and age, reducing their associated systematic uncertainties. We used the GlitchPy code\footnote{\url{https://github.com/kuldeepv89/GlitchPy}} and fitted the observed frequencies for $\mu$ Herculis with the regularisation parameter $\lambda_{\rm A} = 2$ \citep[see Method A presented in the Appendix of][]{verma22}. We reduced $\lambda_{\rm A}$ from 10 to 2 to minimise potential biases in the inferred glitch parameters. The fit to the data is shown in Figure~\ref{fig:glitch fit} and the distributions of fitted parameters obtained using Monte Carlo simulations are shown in Figure~\ref{fig:glitch_realisations}. The inferred values of $\langle A_{\rm He} \rangle$, $\Delta_{\rm He}$, $\tau_{\rm He}$ and $\phi_{\rm He}$ are listed in Table~\ref{tab:table_observed}. Note that Eq.~\ref{eq:2} is not applicable to the mixed modes, and therefore we excluded all $\ell = 1$ modes from the fit because many of them are perturbed due to the presence of mixed modes. Furthermore, we removed all $\ell = 2$ modes with frequencies below 1200 $\mu$Hz from the glitch analysis due to the presence of at least one observed mode below 1200 $\mu$Hz that is likely perturbed by mixed modes (see the \'echelle diagram in Figure~\ref{fig:muHer_echelle}). 

As apparent from Figure~\ref{fig:glitch fit}, we found a relatively large convection zone glitch amplitude, $A_{\rm CZ} = 0.24\pm0.04~\mu$Hz. However, since $\ell = 1$ modes are not included in the fit, and some of $\ell = 2$ modes used may be slightly perturbed due to their possible interaction with the g-modes, the aliasing problem \citep[][]{mazu01} results in a bimodal solution, thus leading to potentially unreliable convection zone glitch parameters. To minimise systematic uncertainties in our inferences, we therefore exclude the convection zone glitch parameters from our modelling procedure.


\section{Stellar modelling}
\label{sec:stellar_modeling}

We used the Garching Stellar Evolution Code \citep[GARSTEC;][]{weiss08} to compute grids of stellar evolution models for their use in stellar modelling. Our reference grid used the following input physics. Nuclear reaction rates were adopted mainly from the NACRE compilation \citep[][]{angu99}, except for $\prescript{14}{}{\rm N}(p, \gamma)\prescript{15}{}{\rm O}$ for which the rate of \citet{form04} was used. We used GARSTEC with the OPAL equation of state \citep{roger02}. The high-temperature OPAL opacities \citep{igle96} were supplemented with the low-temperature values of \cite{ferg05}. The solar metallicity mixture was from \citet[][hereafter AGS09]{aspl09}. We included atomic diffusion following the prescription of \citet{thoul94} and a plane-parallel Eddington grey atmosphere in the models. Since we anticipate a radiative core for $\mu$ Herculis (see G17 and tests below in Section~\ref{sec:ref_grid}), the convective-core overshoot was ignored. Theoretical oscillation frequencies were calculated using the Aarhus adiabatic oscillation package \citep[ADIPLS;][]{chris08}.

In addition to the reference grid, three small grids were calculated to explore the impact of a change in certain input physics on the helium glitch parameters in Section~\ref{sec:freq_glitch_fitting}. We computed these grids, each containing only 10 evolutionary tracks: (1) using the \citet[][hereafter GS98]{grev98} solar metallicity mixture instead of AGS09; (2) using the FreeEOS\footnote{\url{https://freeeos.sourceforge.net/}} equation of state \citep{irwin12} instead of OPAL; and (3) including convective-core overshoot \citep{frey96}. All other input physics in these grids were retained as in the reference setup. We refer the reader to Section~\ref{sec:freq_glitch_fitting} for the usage of these grids and other related details.

\begin{figure}
    \centering
    \includegraphics[width=\linewidth]{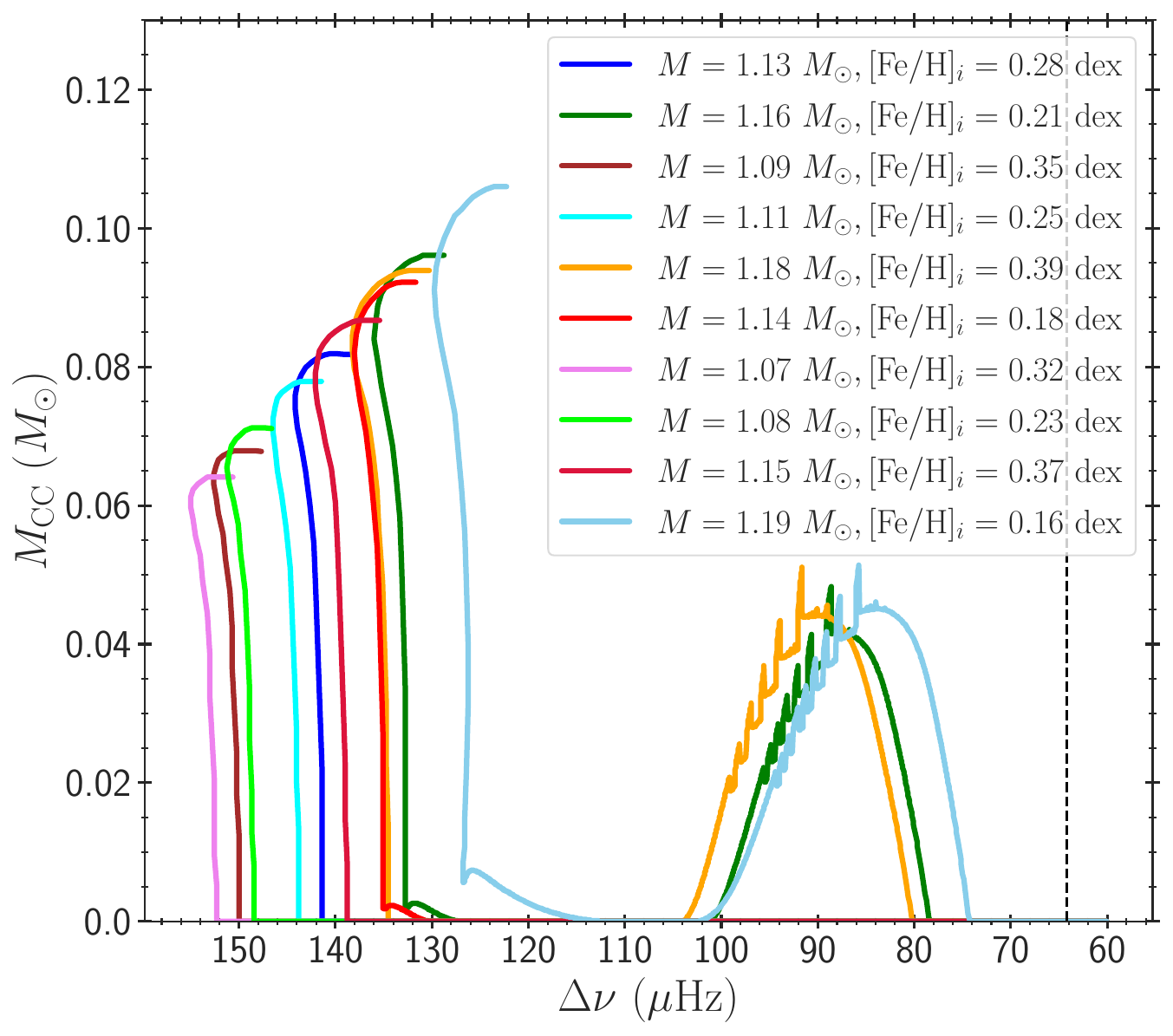}
    \caption{Convective-core mass as a function of large frequency separation for 10 evolutionary tracks uniformly sampled in the mass range $M \in [1.05, 1.20]$ $M_\odot$ and initial metallicity range $[{\rm Fe}/{\rm H}]_i \in [0.14, 0.42]$ dex. The curves represent different tracks with their $M$ and $[{\rm Fe}/{\rm H}]_i$ given in the legend. The initial helium abundance and the mixing length parameter for all the tracks are $0.29$ and $1.85$, respectively. The tracks begin from the zero-age main sequence (ZAMS) and terminate when $\Delta\nu = 60~\mu\mathrm{Hz}$. The dashed vertical line marks the observed large frequency separation of $\mu$ Herculis.}
    \label{fig:convective Overshoot}
\end{figure}


\subsection{Reference model grid}
\label{sec:ref_grid}

Given the observed metallicity and the initial modelling results of G17, we chose the grid parameter space such that the mass $M$ $\in$ [1.05, 1.20] $M_{\odot}$, initial helium abundance $Y_{i}$ $\in$ [0.24, 0.34], initial metallicity $[{\rm Fe}/{\rm H}]_{i}$ $\in$ [0.14, 0.42] dex, and the mixing-length parameter $\alpha_{\rm MLT}$ $\in$ [1.5, 2.2].  
We sampled the above parameter space uniformly using quasi-random numbers \citep{sobol67}. 

Before computing the grid, we performed a few tests to assess the possible impact of ignoring the convective-core overshoot and to decide on the temporal resolution of the grid. Because the hydrogen in the core of subgiants is exhausted, their radiative temperature gradient in the core is below the local adiabatic gradient, making it radiative. However, its ancestors could have had a convective core on the main sequence, and core overshoot can have important implications in that case. In the following, we check whether $\mu$ Herculis had a potentially convective core on the main sequence. Since the presence of convective core depends mainly on the mass and metallicity: the higher the mass and smaller the metallicity, the higher the chance of the presence of a convective core in the star -- we calculate a small test grid with the initial parameter space being $M$ $\in$ [1.05, 1.20] $M_{\odot}$, $[{\rm Fe}/{\rm H}]_{i}$ $\in$ [0.14, 0.42] dex, $Y_{i} $ = 0.29, and $\alpha_{\rm MLT}$ = 1.85. Note that we use central values of the reference grid for $Y_i$ and $\alpha_{\rm MLT}$. We generated 10 uniformly distributed tracks in this parameter space to check the presence of the convective core in the models. In Figure~\ref{fig:convective Overshoot}, we show the convective-core mass, $M_{\rm CC}$, as a function of the large frequency separation, $\Delta\nu$, calculated using the scaling relation \citep{kjel95} for all models in the tracks. The tracks begin on the zero-age main sequence (ZAMS) on the left and continue to the right until $\Delta\nu = 60$ $\mu$Hz. Clearly, none of the tracks has a convective core near the evolutionary state of $\mu$ Herculis (see the observed $\Delta\nu$ in Table~\ref{tab:table_observed}). As we can see, seven tracks do not have a convective core throughout the main sequence, except very close to the ZAMS. The other three tracks have a small convective core; however, they all have masses at the higher end ($> 1.16$ M$_\odot$). Given these results, we chose to exclude the convective-core overshoot in the reference grid (see also Section~{\ref{sec:freq_glitch_fitting}}). 

\begin{figure}

    \centering
    \includegraphics[width=\linewidth]{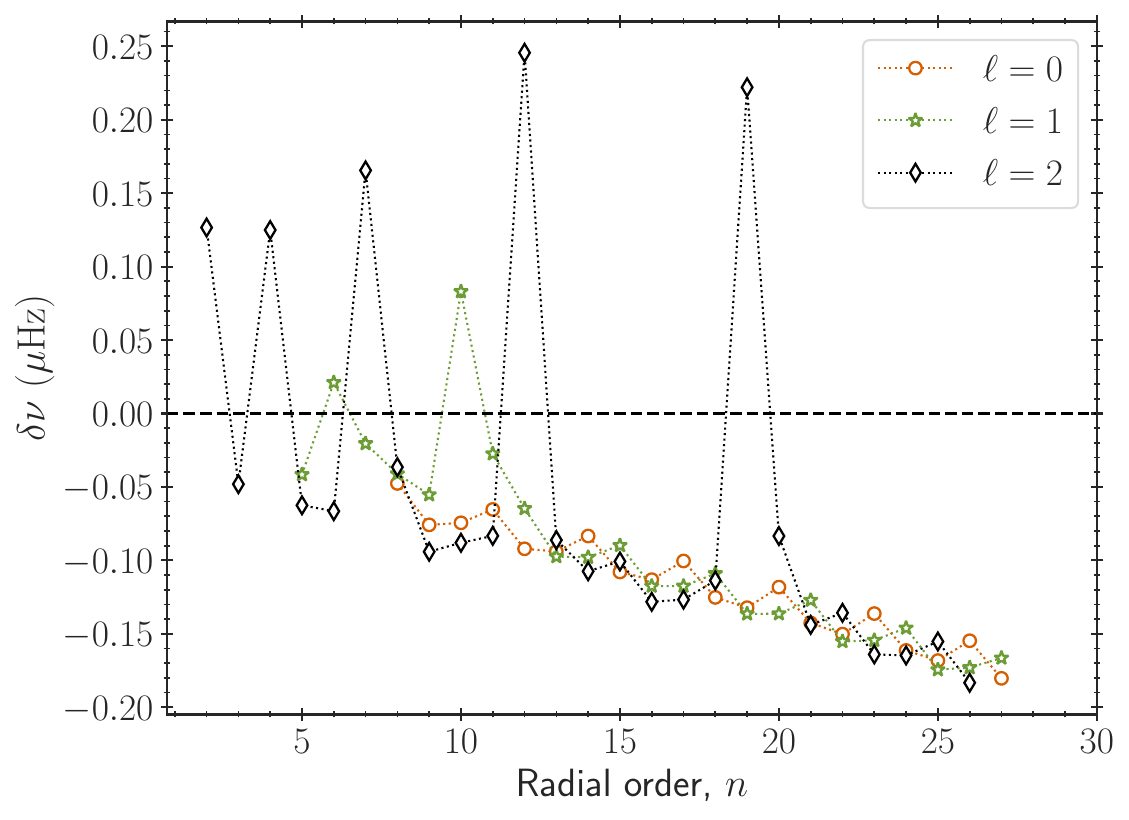}
    \caption{Frequency differences between two consecutive models (``older model" $-$ ``younger model") along an evolutionary track, selected such that their large frequency separation approximates the corresponding observed value of $\mu$ Herculis, plotted as a function of radial order. The track is evolved with stellar mass, initial metallicity, initial helium abundance, and mixing-length parameter of $1.125$ $M_{\odot}$, $0.28$ dex, $0.29$, and $1.85$, respectively. As shown in the legend, the orange circles, green stars, and black diamonds represent the frequency differences for modes with $\ell = 0$, $\ell = 1$, and $\ell = 2$, respectively. The dashed horizontal line marks the zero difference.}
    \label{fig:temporal Resolution}

\end{figure}

It is well known that mixed-mode frequencies observed in subgiants such as $\mu$ Herculis evolve rapidly with time \citep[see e.g.][]{li19}, necessitating a model grid with sufficiently high temporal resolution. However, storing a large number of equilibrium model structures along each evolutionary track to later calculate theoretical frequencies demands considerable disk space. Therefore, we wish to estimate the upper limit on time steps taken during stellar evolution to achieve the desired model frequency resolution while balancing storage constraints. For this, we repeatedly calculated a track representative of $\mu$ Herculis with $M$ = 1.125 $M_{\odot}$, $Y_{i} = 0.29$, $[{\rm Fe}/{\rm H}]_{i} = 0.28$ dex, and $\alpha_{\rm MLT} = 1.85$ (center of the parameter space of the reference grid) with varying upper limit on time steps, and evolved until $\Delta\nu = 60$~$\mu$Hz. We then selected two consecutive models from the track with $\Delta\nu$ values closest to the observed value of $\mu$ Herculis to check the frequency resolution. These models were chosen to ensure adequate temporal resolution at least in the vicinity of the target's evolutionary state. We set the upper limit on time steps such that the frequency differences between consecutive models, for the same set of modes as observed, remain at or below 0.2 $\mu$Hz. This threshold was adopted after visually inspecting the uncertainties on the observed frequencies of $\mu$ Herculis. It should be noted that the helium glitch signature does not evolve as fast as the smooth component of the frequency (or the mixed mode frequencies). In other words, the change in the helium glitch amplitude is much smaller than 0.2 $\mu$Hz. Figure~\ref{fig:temporal Resolution} shows the frequency differences as a function of the radial order for the optimal choice of the upper limit on time steps. The chosen upper limit on time steps leads to a time separation of about 0.26 Myr between the above consecutive models. In the figure, note the larger frequency differences for the mixed quadrupole modes than for the mixed dipole modes. This is expected, as changes in the buoyancy frequency induced by variations in the model affect the $\ell = 2$ mixed-mode frequencies more strongly than the $\ell = 1$ modes, owing to the $\sqrt{\ell(\ell+1)}$ scaling of the g-mode frequencies predicted by asymptotic theory \citep{tass80}. These strongly perturbed mixed-mode quadrupole frequencies have low detectability due to their large mode inertia and are therefore typically excluded during the matching of observed and model frequencies \citep[see e.g.][]{stok19}.

We calculated 3000 tracks distributed uniformly in the parameter space defined at the beginning of this section. Local equilibrium structures were stored only for models with $\Delta\nu$ $\in [60, 70]$ $\mu$Hz, which requires about 5.6 terabytes of disk space. The grid contains about 4.8 million models.


\subsection{Model fitting technique}
\label{sec:fitting_procedure}

We fit the stellar models to the observed data for $\mu$ Herculis using the BAyesian STellar Algorithm \citep[BASTA\footnote{\url{https://basta.readthedocs.io/en/latest/}};][]{agui22,verma22}. BASTA is a versatile Bayesian model fitting tool, which can be used to fit a variety of observables including the effective temperature $T_{\rm eff}$, surface metallicity $[{\rm Fe}/{\rm H}]$, oscillation frequencies $\nu_{nl}$, and the helium glitch properties ($\langle A_{\rm He} \rangle$, $\Delta_{\rm He}$, and $\tau_{\rm He}$). For technical details, we refer the reader to the above references. Briefly, it combines the likelihood of the observed data with prior knowledge of the fitting parameters according to the Bayes theorem to calculate the posterior probability distribution. 

We perform two separate fits: (1) excluding the observed helium glitch properties, i.e. fitting only $\nu_{nl}$ together with $T_{\rm eff}$ and $[{\rm Fe}/{\rm H}]$ (Fit1); and (2) including $\langle A_{\rm He} \rangle$, $\Delta_{\rm He}$, and $\tau_{\rm He}$ as well (Fit2). We define the likelihood function as $\mathcal{L} = e^{-\chi^{2}/2}$. In the case of Fit1, $\chi^{2}$ in the expression for $\mathcal{L}$ is defined by:
\begin{equation}
    \chi^{2}_{1} = \left(\frac{T_{\rm eff}^{\rm o}  -  T_{\rm eff}^{\rm m}}{\sigma_{T_{\rm eff}}}\right)^2 + \left(\frac{[{\rm Fe}/{\rm H}]^{\rm o}  -  [{\rm Fe}/{\rm H}]^{\rm m}}{\sigma_{[{\rm Fe}/{\rm H}]}}\right)^2 + \frac{1}{N}\sum_{nl}\left(\frac{\nu_{nl}^{\rm o}  - \nu_{nl}^{\rm m}}{\sigma_{nl}}\right)^2,
\end{equation}  
where $N = 48$ is the number of observed modes used. The superscripts `o' and `m' refer to the observation and model, respectively. The model frequencies were corrected for the surface effect following the two-term prescription of \citet{ball14}. To mitigate the residual systematic uncertainties in the model frequencies, the last term was weighted by a factor of $1/N$ \citep[see][]{cunh21}. In the case of Fit2, $\chi^{2}$ in the expression for $\mathcal{L}$ includes an additional term, given by:
\begin{equation}
\label{eq:5}
    \chi^{2}_{\rm g} = \frac{1}{3}(\textbf{g}^{\rm o} - \textbf{g}^{\rm m})^{\rm T}\textbf{C}^{-1}(\textbf{g}^{\rm o} - \textbf{g}^{\rm m}),
\end{equation}
where $\textbf{g}$ = ($\langle A_{\rm He} \rangle$, $\Delta_{\rm He}$, $\tau_{\rm He}$) is a vector containing the three helium glitch observables and \textbf{C} is its covariance matrix calculated using the GlitchPy code. The superscript `T' means the transpose of the column vector. Again, we choose to weight $\chi^{2}_{\rm g}$ by the number of observables, resulting in the factor $1/3$. Therefore, the total chi-square in this case is given by:
\begin{equation}
    \chi^{2}_{\rm 2} = \chi^{2}_{\rm 1} + \chi^{2}_{\rm g}.
\end{equation}
Note that we consistently used the same set of model modes as for the observed data, together with the observed frequency errors, in the glitch fitting process to infer the model values of $\langle A_{\rm He} \rangle$, $\Delta_{\rm He}$, and $\tau_{\rm He}$. The glitch analysis involves nonlinear optimisation in a high dimensional space and is computationally expensive. To improve computational efficiency, we evaluate the likelihood only for models whose $T_{\rm eff}$ and $[{\rm Fe}/{\rm H}]$ agree with the observed values within $3\sigma$ and 0.25 dex, respectively, and whose frequency corresponding to the lowest frequency observed radial mode, $\nu_{\rm min}^o$, lies in [$\nu_{\rm min}^o - 3\sigma_{\rm min}$, $\nu_{\rm min}^o + 0.15\Delta\nu$], where $\sigma_{\rm min}$ is the observational uncertainty on $\nu_{\rm min}^o$. The asymmetry in the frequency range around $\nu_{\rm min}^{o}$ accounts for the surface effect.

We used the initial mass function \citep{salp55} as prior on the mass. The posterior probability was used to derive the stellar parameters and associated uncertainties following \citet{agui22}.


\section{Results}
\label{sec:results}

In this section, we present the results of the modelling of our subgiant target $\mu$ Herculis. First, we show the results obtained by fitting the individual oscillation frequencies only. 
Later, we include additional constraints from the helium glitch analysis and present the outcome. Finally, we discuss the origin of the helium glitch signature and provide a physical interpretation of our findings for $\mu$ Herculis. 

\begin{table} 

    \caption{Observable parameters of the best-fitting models M1 and M2, obtained using the two fitting approaches, Fit1 and Fit2 (see text for details).}
    \label{tab:table_models}
    \centering
    \begin{tabularx}{\linewidth}{>{\centering\arraybackslash}X|>{\centering\arraybackslash}X|>{\centering\arraybackslash}X|>{\centering\arraybackslash}X|>{\centering\arraybackslash}X|>{\centering\arraybackslash}X|>
    {\centering\arraybackslash}X|>
    {\centering\arraybackslash}X|>
    {\centering\arraybackslash}X}
    
        \hline
        \hline
        \noalign{\vskip 0.1cm}    

        \hspace*{-0.1cm}Model & 
        \makebox[0.1pt][l]{\hspace*{-0.35cm}[Fe/H]} & 
        $T_{\rm eff}$ & 
        $R$ & 
        $L$ & \makebox[-0.1cm][c]{$\langle A_{\rm He} \rangle$} & $\Delta_{\rm He}$ & $\tau_{\rm He}$ & $\phi_{\rm He}$\\
        
         & 
        \makebox[0.1pt][l]{\hspace*{-0.25cm}(dex)} & 
        (K) & 
        ($R_\odot$) & 
        ($L_\odot$) & \makebox[-0.005cm][c]{ ($\mu$Hz)}& (s) & (s) & \\[0.5ex]
        
        \noalign{\vskip 0.1cm}
        \hline 
        \noalign{\vskip 0.1cm}

        M1 & $0.21$ & $5572$ & $1.73$ & $2.59$ & $0.37$ & $166$ & $1593$ & $2.66$\\
        \noalign{\vskip 0.1cm}
        \hline 
        \noalign{\vskip 0.1cm}
        
        M2 & $0.25$ & $5582$ & $1.71$ & $2.54$ & $0.43$ & $161$ & $1560$ & $2.96$\\
        \noalign{\vskip 0.1cm}
        \hline 
              
    \end{tabularx}

\end{table}

\begin{figure*}

    \centering
    \includegraphics[width=0.8                  \linewidth]{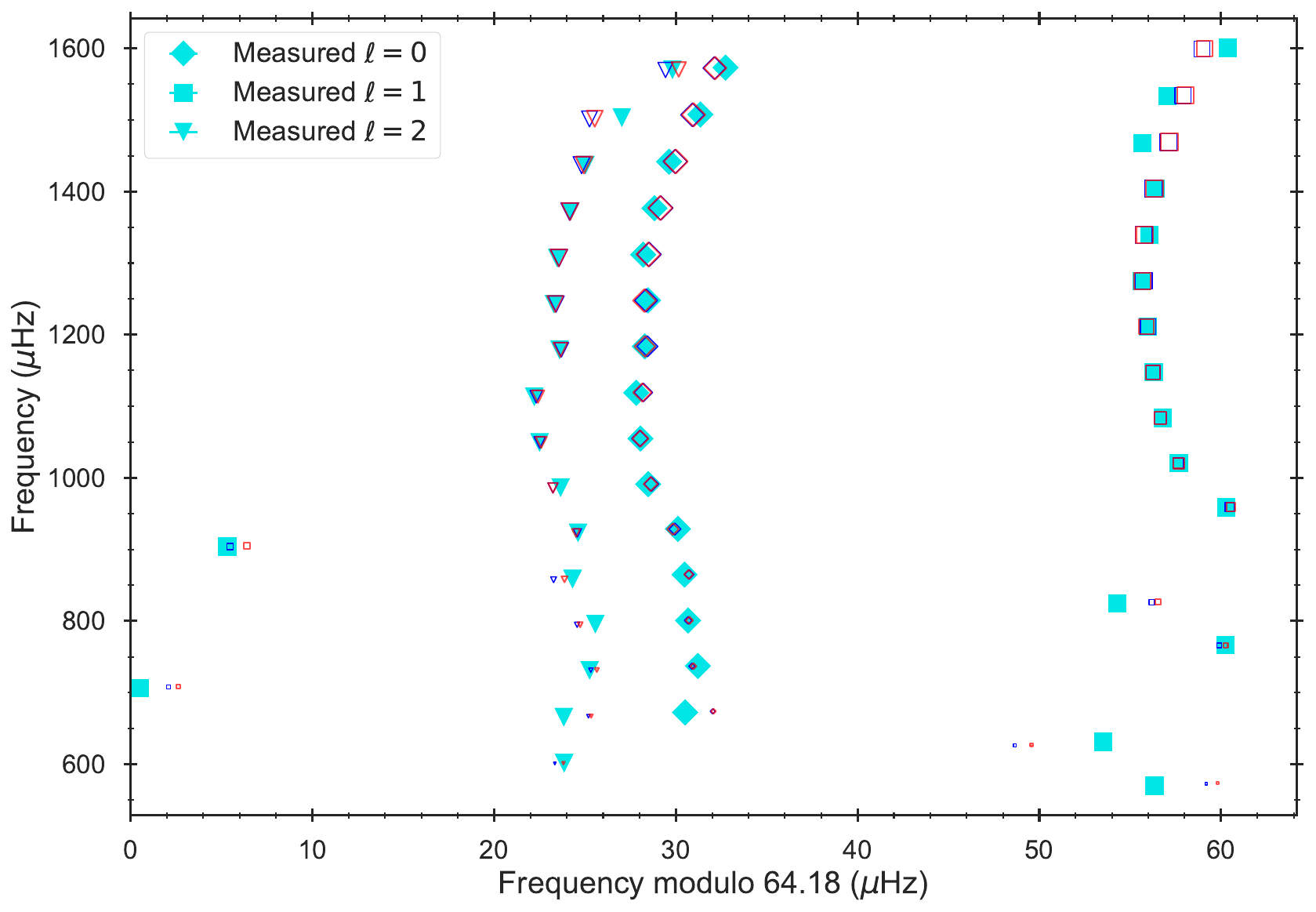}
    \caption{Comparison of the observed and best-fitting model frequencies in an \'echelle diagram. The filled, cyan diamonds, squares, and triangles represent the observed oscillation modes with harmonic degrees $\ell = 0$, $\ell = 1$, and $\ell = 2$, respectively. The corresponding empty, red and blue symbols represent oscillation frequencies of the best-fitting models M1 and M2 obtained from fitting approaches Fit1 and Fit2, respectively (see the text). The model symbol sizes are scaled inversely with their normalized mode inertias.}
    \label{fig:muHer_echelle} 
    
\end{figure*}

\subsection{Inferences based on frequencies only (Fit1)}
\label{sec:freq_fitting}

As mentioned in Section~\ref{sec:fitting_procedure}, we first fit $T_{\rm eff}$, $[{\rm Fe}/{\rm H}]$ and $\nu_{nl}$ using the BASTA package. The resulting posterior probability distribution is shown in Figure~\ref{fig:corner_plot_freqs}. For the best-fitting model (referred to hereafter as M1), the effective temperature, surface metallicity, luminosity, radius, and the helium glitch parameters are listed in the first row of Table~\ref{tab:table_models}. The model not only fits the observed $[{\rm Fe}/{\rm H}]$ and $T_{\rm eff}$ in Table~\ref{tab:table_observed} within 1$\sigma$ but also the observed $R$ and $L$ values. However, it does not fit the observed helium glitch parameters well. In an \'echelle diagram shown in Figure~\ref{fig:muHer_echelle}, we compare the surface-corrected best-fitting model frequencies with the observed ones. As we can see, both the p-mode and the mixed-mode observed frequencies are reasonably well reproduced by the model. The small differences in the mixed-mode model and observed frequencies could be partly due to limited grid resolution and partly because of shortcomings in the stellar evolution models (see Section~\ref{sec:freq_glitch_fitting}). 

The inferred stellar parameter values along with associated statistical uncertainties are listed in the first row of Table~\ref{tab:table_median}. The asymmetric error bars on some of the parameters primarily arise from the limited resolution of the model grid. Although our grid is very dense compared to grids typically used for main-sequence stars -- especially considering the relatively small parameter space -- it is only just sufficient to adequately sample the posterior distribution for this subgiant (see Figure~\ref{fig:corner_plot_freqs}). As we can see in Table~\ref{tab:table_median}, the inferred $M$ and $R$ are in excellent agreement with the values reported in G17 (see the last row in Table~\ref{tab:table_median}). However, our age is significantly older (by 2$\sigma$) than that of G17. Note also that our $Y_s = 0.224^{+0.024}_{-0.008}$ and $Y_i$ in the table are subsolar \citep[for the Sun, $Y_s = 0.2485\pm0.0035$ and $Y_i = 0.278\pm0.006$;][]{basu04b,sere10} despite the significantly supersolar metallicity of $\mu$ Herculis. As we shall see in the next section, including the helium glitch constraints in the fit alleviates these discrepancies to some extent.


\subsection{Inferences based on helium glitch constraints (Fit2)}
\label{sec:freq_glitch_fitting}

\begin{figure}
    \centering
    \includegraphics[width=\linewidth]{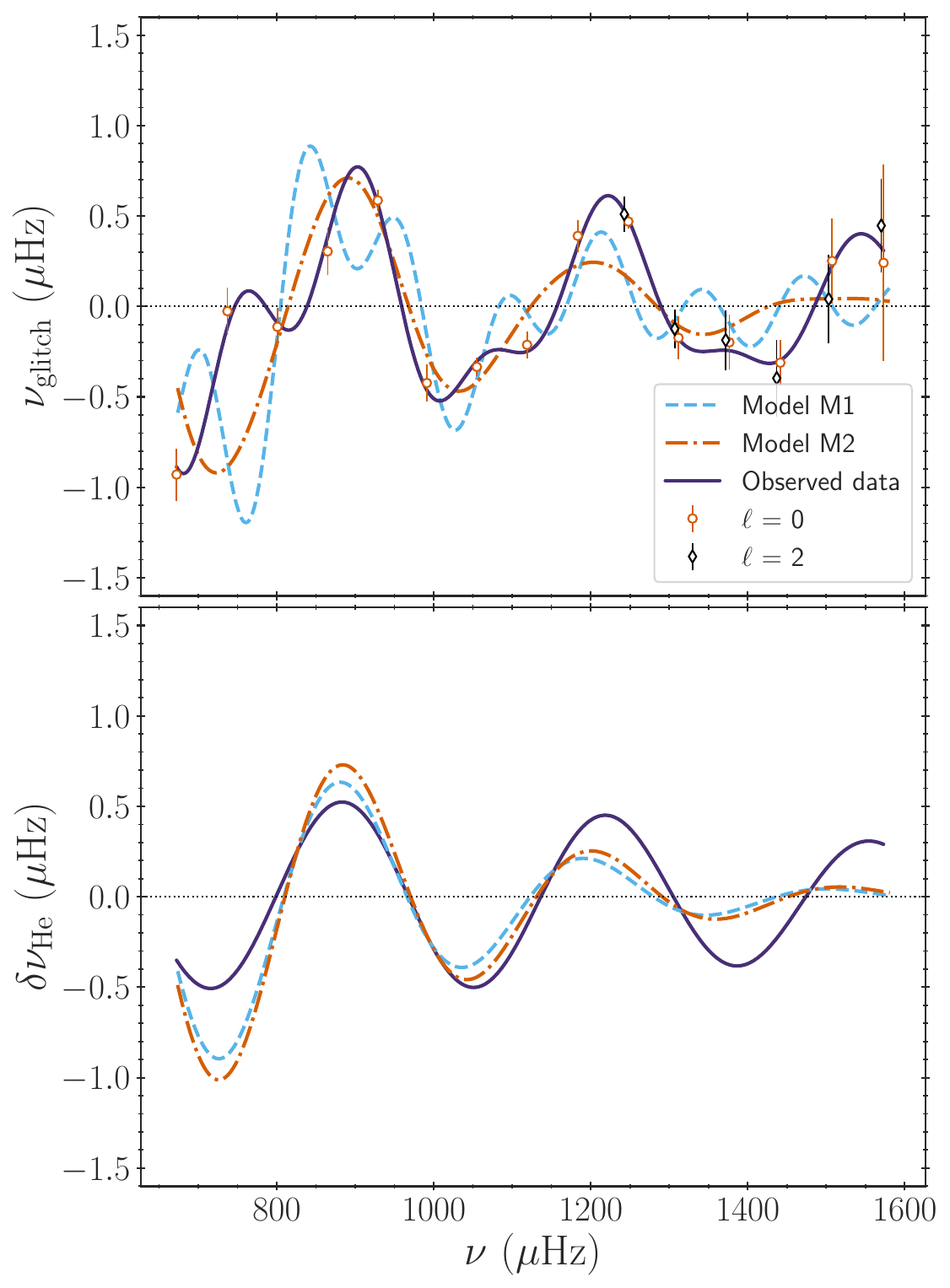}
    \caption{Comparison of the observed and best-fitting model glitch signatures. The top panel shows the sum of glitch signatures arising from the helium ionisation zone and the base of the envelope convection zone for the observed data and the best-fitting models obtained using the approaches Fit1 (model M1) and Fit2 (model M2). The circles and diamonds represent the observed $\ell = 0$ and $\ell = 2$ modes, respectively, and the solid curve shows their best-fit. The dashed and dot-dashed curves represent fits to the frequencies of models M1 and M2, respectively (individual model frequencies are not shown in the figure for a cleaner presentation). The bottom panel shows the fitted helium glitch signatures (see Eq.~\ref{eq:2}) for the observed data and best-fitting models, M1 and M2. The dotted horizontal lines in both panels indicate the zero level.}
    \label{fig:glitch_panel}
\end{figure}

\begin{table*} 

    \caption{Inferred stellar parameters from the two fitting approaches, Fit1 and Fit2 (see text for details). The last row includes values reported by G17, when available.}
    \label{tab:table_median}
    \centering
    \begin{tabularx}{\linewidth}{>{\centering\arraybackslash}X|>{\centering\arraybackslash}X|>{\centering\arraybackslash}X|>{\centering\arraybackslash}X|>{\centering\arraybackslash}X|>{\centering\arraybackslash}X|>
    {\centering\arraybackslash}X|>
    {\centering\arraybackslash}X|>
    {\centering\arraybackslash}X}
    
        \hline
        \hline
        \noalign{\vskip 0.1cm}
        
        Technique & 
        Mass & Radius & $[{\rm Fe}/{\rm H}]_{i}$ & $Y_{i}$ &
        $Z_{i}$ &
        $\alpha_{\rm MLT}$ &
        Age & Chi-square\\
         & ($M_{\odot}$) & ($R_{\odot}$) & (dex) &  &  & & (Gyr) \\
        \noalign{\vskip 0.1cm}
        
        \hline
        
        \noalign{\vskip 0.1cm}
        Fit1 & $1.117^{+0.040}_{-0.043}$ &
        $1.714^{+0.022}_{-0.021}$ & 
        $0.289^{+0.046}_{-0.073}$ &
        $0.263^{+0.030}_{-0.004}$ &
        $0.025^{+0.002}_{-0.004}$ &
        $1.80^{+0.07}_{-0.14}$ &
        $8.8^{+0.6}_{-0.4}$ & 49.2\\

        \noalign{\vskip 0.1cm}
        
        \hline

        \noalign{\vskip 0.1cm}
        Fit2 &
        $1.105^{+0.058}_{-0.024}$ &
        $1.709^{+0.030}_{-0.015}$ &
        $0.320^{+0.014}_{-0.051}$ &
        $0.290^{+0.004}_{-0.028}$ &
        $0.027^{+0.001}_{-0.003}$ &
        $1.80^{+0.09}_{-0.11}$ &
        $8.4^{+0.4}_{-0.1}$ & 58.1\\

        \noalign{\vskip 0.1cm}

        \hline
        \noalign{\vskip 0.1cm}

        G17 &        
        $1.11^{+0.01}_{-0.01}$ & 
        $1.71^{+0.01}_{-0.02}$ &
        - &
        - & - & - &
        $7.8^{+0.3}_{-0.4}$ & -\\
        
        \noalign{\vskip 0.1cm}          
        \hline
        
    \end{tabularx}

\end{table*}

\begin{figure}

    \centering
    \includegraphics[width=1.0\linewidth]{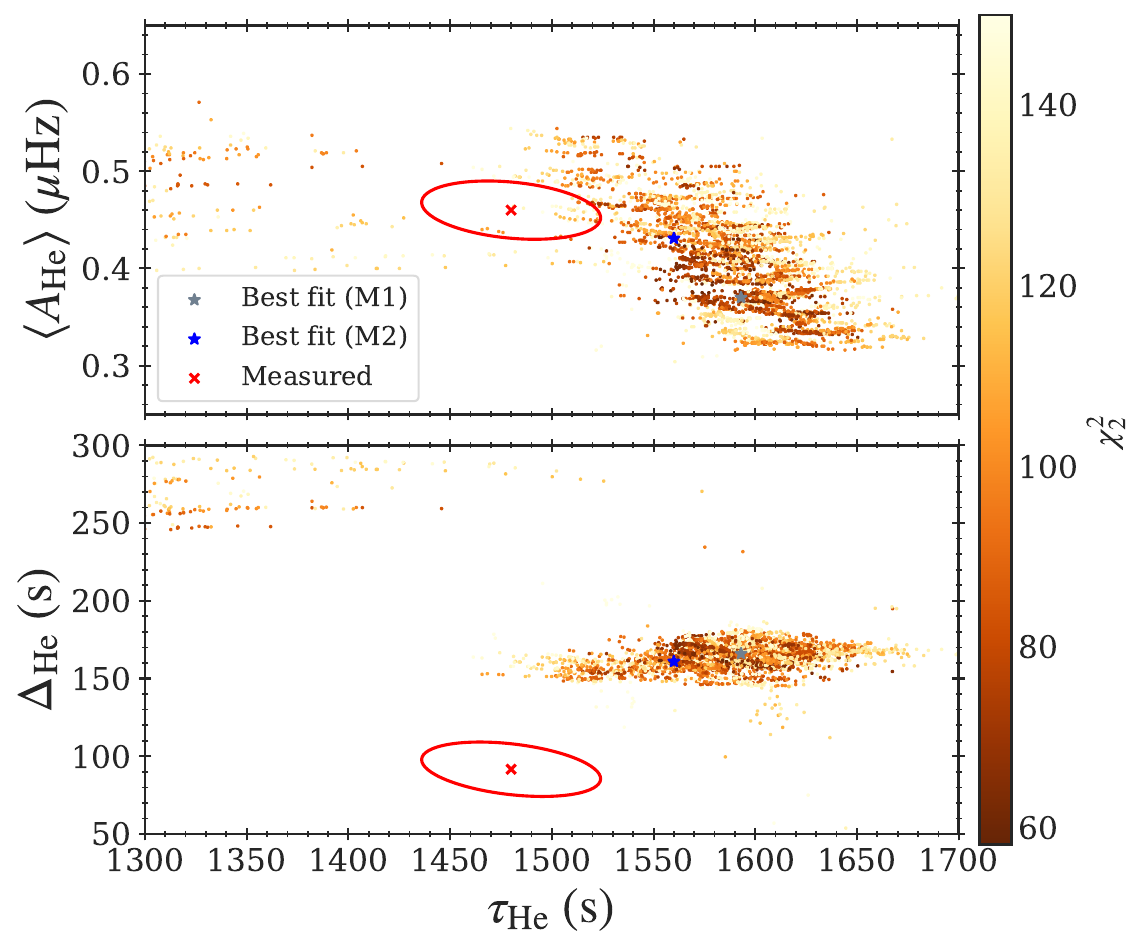}
    \caption{Average amplitude of helium signature (top panel) and acoustic width of helium ionisation zone (bottom panel) as a function of the acoustic depth of the helium ionisation zone. The red cross in each panel indicates the measured helium glitch properties with the red ellipse highlighting a 1$\sigma$ confidence region. The dots in each panel represent the glitch parameters for all models with $\chi_{2}^2 < 150$. The dots are color-coded by their $\chi_{2}^2$ values. The grey and blue star symbols in each panel represent the best-fitting models M1 and M2, respectively.}
    \label{fig:glitch parameters}
    
\end{figure}

We now fit the helium glitch parameters in addition to the observables fitted in the previous section. The resulting posterior probability distribution is shown in Figure~\ref{fig:corner_plot_glitch}. For the best-fitting model (hereafter referred to as M2), all relevant parameters are listed in the second row of Table~\ref{tab:table_models}. Again, $[{\rm Fe}/{\rm H}]$, $T_{\rm eff}$, $R$, and $L$ are within 1$\sigma$ of the corresponding observed values in Table~\ref{tab:table_observed}. In Figure~\ref{fig:glitch_panel}, we compare the observed glitch signatures of $\mu$ Herculis with those of the best-fitting models M1 and M2. In the top panel, the relatively high-frequency, low-amplitude modulation arising from the base of the convection zone is typically not robust, primarily due to its weak signature and the problem of aliasing \citep{mazu01,verma17}. This makes the fits appear significantly different. In contrast, the fit to the high-amplitude helium glitch signature, as shown in the bottom panel, is reliable. As expected, the helium glitch parameters for model M2 in Table~\ref{tab:table_models} fit the corresponding observed data better than model M1. In particular, the $\langle A_{\rm He} \rangle$ value of model M2 lies within 1$\sigma$ of the observed value, in contrast to M1, for which it deviates by 3$\sigma$. However, a discrepancy exceeding the 3$\sigma$ level between the observed and modeled $\Delta_{\rm He}$ still persists, which we further investigate later in this section. Model M2 also fits the observed data well in the \'echelle diagram (see Figure~\ref{fig:muHer_echelle}). We note that neither the track associated with model M1 nor that of M2 exhibits a convective core during the main-sequence phase, which is consistent with our choice to exclude convective-core overshoot. 

\begin{figure*}

    \centering
    \includegraphics[width=0.8\linewidth]{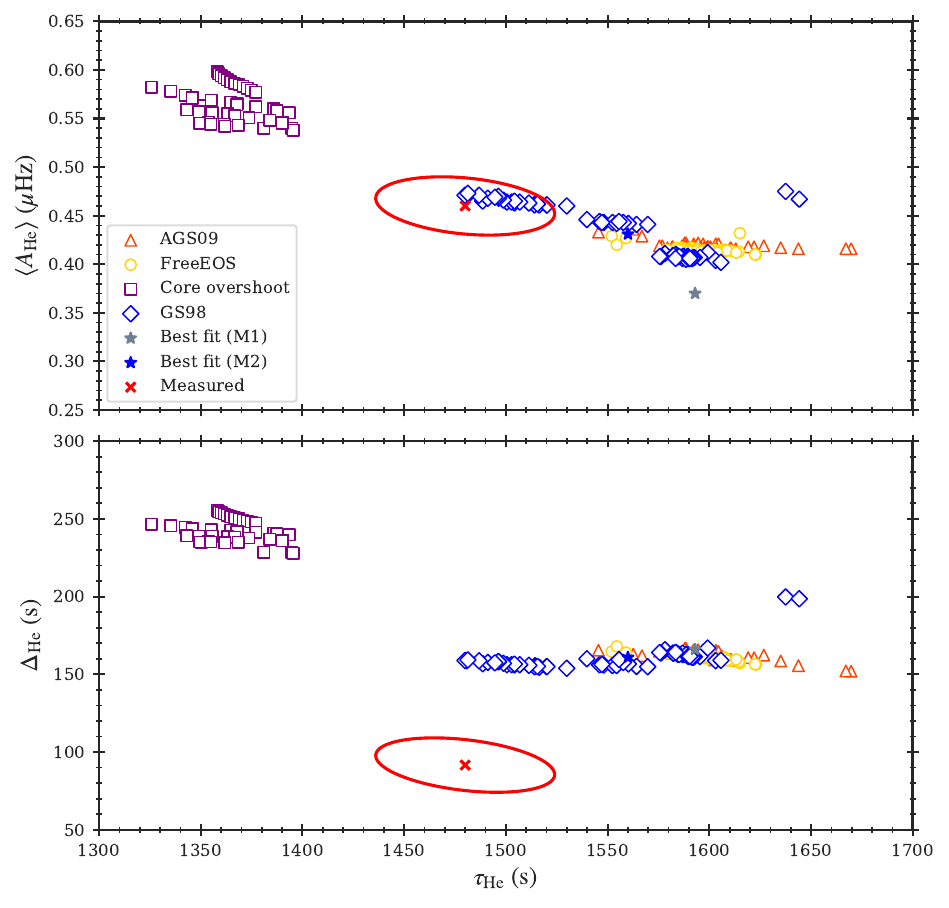}
    \caption{Average amplitude of helium signature (top panel) and acoustic width of helium ionisation zone (bottom panel) as a function of the acoustic depth of the helium ionisation zone. The red cross in each panel indicates the measured helium glitch properties with the red ellipse highlighting a 1$\sigma$ confidence region. The orange triangles, blue diamonds, yellow circles, and purple squares represent models with $\chi_{2}^2 < 150$ obtained using the AGS09 solar metallicity mixture (corresponds to the reference input physics used in this study), GS98 mixture, FreeEOS, and the core overshoot of $f_{\rm OV} = 0.015$, respectively (see text for details). For comparison, the best-fitting models M1 and M2 are shown in each panel as grey and blue stars, respectively.}
    \label{fig:solar_mixtures_comparision}
    
\end{figure*}

The inferred values of the stellar parameters, together with associated statistical uncertainties, are given in the second row of Table~\ref{tab:table_median}. The inferred $M$ and $R$ from Fit2 are again in excellent agreement with the values obtained from Fit1 and those reported in G17. As expected, the large frequency separations for the best-fitting models M1 and M2 are nearly identical, being $64.12~\mu$Hz and $64.09~\mu$Hz, respectively. It is interesting to note that the age has been reduced, but it is still consistent with the value obtained from Fit1 and is significantly higher than the G17 value. The inferred value of $Y_{s} = 0.242^{+0.006}_{-0.021}$ and the initial helium abundance $Y_i$ listed in the table are both higher than those obtained with Fit1, as expected due to the supersolar metallicity of $\mu$ Herculis. Furthermore, since model M2 reproduces the average amplitude more accurately than model M1, we believe that the $Y_s$ and $Y_i$ -- and consequently the age -- derived from Fit2 are more accurate. It is interesting to note from Figures~\ref{fig:corner_plot_freqs} and \ref{fig:corner_plot_glitch}, particularly in the $Y_i$ panels, how the inclusion of the helium glitch parameters in the fit influences the posterior probability of different models.

We now take a closer look at the model fit to the observed helium glitch parameters. Figure~\ref{fig:glitch parameters} shows the helium glitch parameters for all models with $\chi_{2}^2 < 150$. Moreover, it highlights the best-fitting model M2 and displays the observed helium glitch parameters, accompanied by their associated 1$\sigma$ confidence ellipses. Clearly, the ellipses are further away from the cluster of model points, highlighting a significant discrepancy between the model predictions and the observation. The best-fitting model M2 shows marginal discrepancy from the observed value of $\tau_{\rm He}$ ($< 2\sigma$), while the discrepancy for $\Delta_{\rm He}$ is more pronounced ($> 3\sigma$). 

\begin{figure*}

    \centering
    \includegraphics[width=1.0\linewidth]{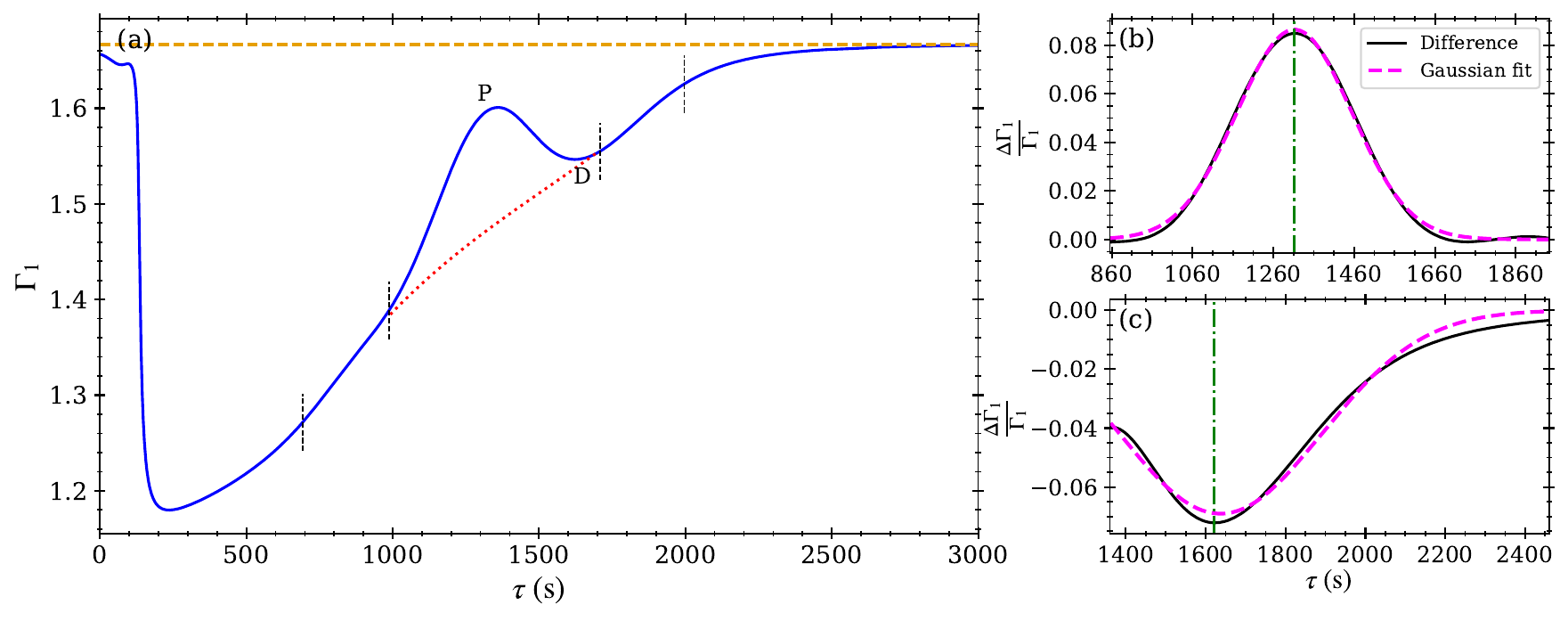}
    \caption{First adiabatic index as a function of acoustic depth for the best-fitting model M2. (a) The blue curve shows $\Gamma_{1}$ profile across the hydrogen and helium ionization zones. The peak and the dip in the profile are labeled by P and D. The dashed orange line at $\Gamma_{1} \approx 1.67$ shows the conventional smooth background. The dotted red curve represents the smooth background, obtained by fitting a cubic polynomial to the segments of the $\Gamma_{1}$ profile on either side of the peak, bounded by the pair of dashed vertical lines. (b) The black curve shows the relative difference between the $\Gamma_1$ profile and the dotted background in the vicinity of the peak. The location of the peak is marked by the green vertical line. The dashed pink curve is a Gaussian fit to the black curve. (c) The black curve shows the relative difference between the $\Gamma_1$ profile and the dashed background in the vicinity of the dip. The location of the dip is marked by the green vertical line. The left-most point in this panel corresponds to the point P. The dashed pink curve is a Gaussian fit to the black curve.}
    \label{fig:gamma_1 profile}
    
\end{figure*}

To assess the influence of the adopted input physics on the observed discrepancies, we computed three small model grids in the vicinity of our best-fitting Model M2, as described in Section~\ref{sec:stellar_modeling}. Since the helium glitch properties may depend to some extent on metallicity and the equation of state (EOS), we computed two grids: one by adopting the solar metallicity mixture of GS98 in place of the reference AGS09, and the other by replacing the OPAL equation of state (EOS) with FreeEOS. A third grid was calculated by including the convective-core overshoot with $f_{\rm OV} = 0.015$ \citep{frey96}. We calculated 10 tracks for each grid in a parameter space constituted by $M$ $\in$ [1.08, 1.14] $M_{\odot}$, $Y_{i}$ $\in$ [0.26, 0.32], $[{\rm Fe}/{\rm H}]_{i}$ $\in$ [0.27, 0.37] dex, and $\alpha_{\rm MLT}$ $\in$ [1.6, 2.0]. This parameter space was uniformly sampled using quasi-random numbers. The models with $\Delta\nu \in [60, 70]$ $\mu$Hz in each track were stored. For a fair comparison, we also computed a fourth grid within the same parameter space, using the same input physics as that of the reference grid.
 
Figure \ref{fig:solar_mixtures_comparision} shows the results of the four model grids. We show all models with $\chi_{2}^2 < 150$. The results obtained from the grid employing FreeEOS are qualitatively consistent with those from the reference grid, suggesting that the observed discrepancies are unlikely to arise from the choice of the EOS. It is interesting to note that the grid using the GS98 solar metallicity mixture yields predictions for $\langle A_{\rm He} \rangle$ and $\tau_{\rm He}$ for some models that are consistent with the observed values within $1\sigma$; however, the discrepancy in $\Delta_{\rm He}$ persists, unfortunately. This indicates that the models with GS98 metallicity mixture may yield a better fit to the helium glitch signature observed in $\mu$ Herculis than the models with AGS09 mixture, potentially related to the corresponding preference for GS98 (which has higher metallicity than AGS09) relative to AGS09 in solar modelling \citep[][]{basu08,serenelli09}. Finally, predictions of the model grid with core overshoot, $f_{\rm OV} = 0.015$, fail to reproduce any of the helium glitch parameters. All such models show $\tau_{\rm He}$ values that are too small and $\langle A_{\rm He} \rangle$ and $\Delta_{\rm He}$ values that are too large, similar to the group of models on the left in Figure \ref{fig:glitch parameters} with on average high chi-square. The substantial change in helium glitch parameters with core overshoot may be an indirect result of its impact on the stellar mass, contrary to our expectation that core overshoot would be unimportant for modelling this star. We wish to emphasise here that these results were obtained from very small, low-resolution grids centered around model M2, and should therefore be interpreted with caution. Nevertheless, these grids are adequate for our primary aim of assessing the influence of input physics on the helium glitch properties.


\subsection{Origin and interpretation of the helium glitch}
\label{sec:location}

Traditionally, it is assumed that the localised depression in the first adiabatic index in the second helium ionisation region causes a glitch in the acoustic structure of solar-type stars and leads to an observable oscillatory signature in the oscillation frequency as a function of radial order. In the left panel of Figure~\ref{fig:gamma_1 profile}, we show the $\Gamma_1$ depression relative to the dashed horizontal background for the best-fitting model M2. \citet{houd07} assumed a Gaussian functional form (see Eq.~\ref{eq:1}) for this depression and derived the corresponding expression for the helium glitch signature in the stellar oscillation frequencies (see Eq.~\ref{eq:2}). If the helium signature observed in the frequencies indeed originates from the second helium ionisation zone, then the value of $\tau_{\rm He}$ obtained from the glitch analysis should correspond to the acoustic depth of a layer where $\Gamma_1$ reaches its minimum (denoted as point D in Figure~\ref{fig:gamma_1 profile}). Furthermore, the value of $\Delta_{\rm He}$ should correspond to the acoustic width of the $\Gamma_1$ depression. While investigating the prospects of asteroseismic inference of the helium abundance in red-giant stars, \citet{broom14} analysed theoretical oscillation frequencies of red-giant models and surprisingly found that the resulting $\tau_{\rm He}$ represents the acoustic depth of a layer close to a point P in Figure~\ref{fig:gamma_1 profile} where $\Gamma_1$ attains its maximum. In a detailed theoretical study, \citet{verma14b} found similar results for main-sequence stars. In the following, we shall carefully examine the origin of the observed helium glitch signature in our subgiant target, $\mu$ Herculis. 

\begin{table*} 
    \caption{Acoustic depths and widths of the helium ionisation zone for the best-fitting models M1 and M2, derived from their oscillation frequencies and sound speed profiles. The columns ``Fitted” list values obtained by fitting the helium glitch signatures in frequencies, i.e. $\tau_{\rm He}$ and $\Delta_{\rm He}$, while ``Actual (peak)” and ``Actual (dip)” give the properties of the $\Delta\Gamma_1/\Gamma_1$ peak and dip from the sound speed profiles, respectively. Depths from the sound speed profile are measured relative to the acoustic surface (see text for details).}
    \label{tab:table_sound}
    \centering
    \begin{tabularx}{\linewidth}{
        >{\centering\arraybackslash}X|
        >{\centering\arraybackslash}X
        >{\centering\arraybackslash}X
        >{\centering\arraybackslash}X|
        >{\centering\arraybackslash}X
        >{\centering\arraybackslash}X
        >{\centering\arraybackslash}X
    }
        \hline\hline
        \noalign{\vskip 0.1cm} 
        \raisebox{-0.25cm}{Parameters}\vspace{-0.4cm} & \multicolumn{3}{c|}{Model M1} & \multicolumn{3}{c}{Model M2} \\
        \noalign{\vskip 0.1cm}
        \cline{2-7}
        \noalign{\vskip 0.1cm}
        
        &Fitted& \hspace{-0.2cm} Actual (peak) & Actual (dip) &  Fitted& \hspace{-0.2cm} Actual (peak) & Actual (dip)\\
        \noalign{\vskip 0.1cm}
        \hline
        \noalign{\vskip 0.1cm}

        Acoustic depth (s)& 1593 & 1619 & 1935 & 1560 & 1600 & 1910\\

        \noalign{\vskip 0.1cm}
        \hline
        \noalign{\vskip 0.1cm}
        
        Acoustic width (s)& 166 & 146 & 252 & 161 & 142 & 254\\
        \noalign{\vskip 0.1cm}
        \hline
        \noalign{\vskip 0.1cm}
        \end{tabularx}
\end{table*}

\begin{figure}
    \centering
    \includegraphics[width=1.0\linewidth]{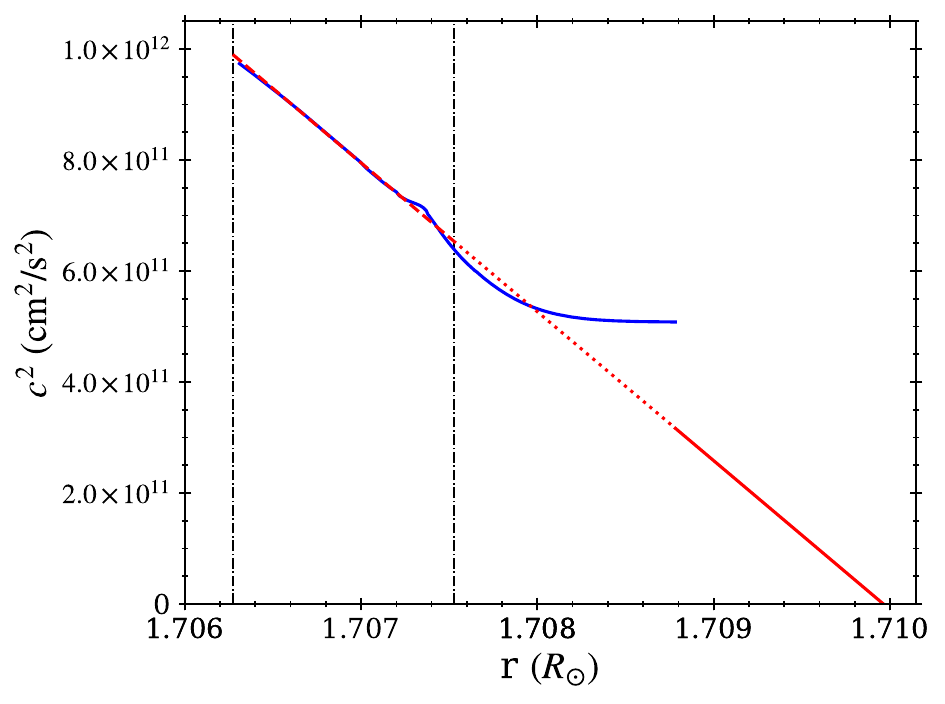}
    \caption{Squared sound speed (blue curve) as a function of the radial coordinate for the best-fitting model M2. The right vertical line indicates the outermost convective layer, while the left vertical line marks a point within the envelope convection zone, located at a distance from the right line equal to that between the right line and the outermost mesh point. The red dashed line is the fit to the $c^2$ segment between the two vertical lines. The red dotted line between the outermost convective layer and the outermost mesh point and the red continuous line between the outermost mesh point and the acoustic surface are the extension of the fitted red dashed line.}
    \label{fig:c2_muHer}
    
\end{figure}

Given the local sound speed $c$ of a stellar model, the acoustic depth $\tau$ of a layer located at radial coordinate $r$ can be calculated using the expression,
\begin{equation}
\label{eq:acoustic_depth}
    \tau = \int_{r}^{R_*} \frac{dr}{c},
\end{equation}
where $R_*$ denotes the radius of the acoustic surface. Note that the stellar photosphere and the acoustic surface are not the same. Since the squared sound speed decreases approximately linearly with $r$ in the outer layers of the envelope convection zone, \citet{balm90} defined the acoustic surface as an atmospheric layer in which the extrapolated $c^2$ from the outer convective layers vanishes. For model M2, we identified its acoustic surface using this definition (see Figure~\ref{fig:c2_muHer}) and estimated the acoustic depth of the photosphere, which was found to be approximately 394 s. Note that this value can vary slightly (by a few tens of seconds) depending on the exact extent of the outer convective layers used to fit the straight line. As expected for a subgiant, the above value for $\mu$ Herculis is larger than the corresponding value for the Sun \citep[225 s;][]{houd07}.

\begin{figure}

    \centering
    \includegraphics[width=1.0\linewidth]{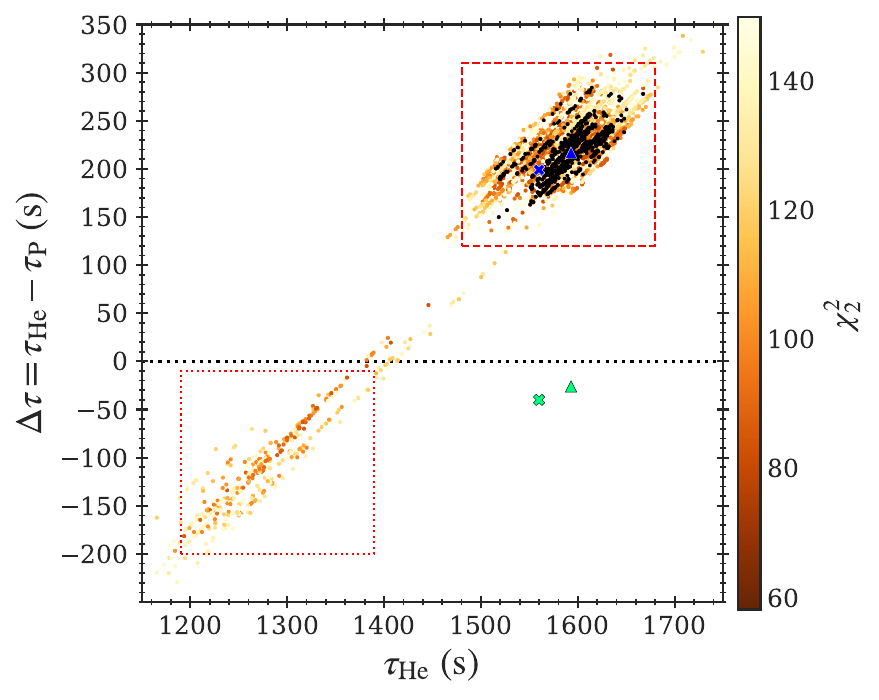}
        \caption{Difference between the acoustic depth $\tau_{\rm He}$ inferred from the glitch analysis and the acoustic depth of the $\Gamma_{1}$ peak, $\tau_{\rm P}$, calculated from the outermost mesh point using the sound speed profile as a function of $\tau_{\rm He}$. The dots, color-coded by $\chi_{2}^2$, represent all models with $\chi_{2}^2 < 150$, while the black dots indicate models with $\chi_{2}^2 < 80$. The blue triangle and cross show the (uncorrected) best-fitting models M1 and M2, respectively, while the corresponding green symbols represent the corrected ones (see text for details). Two groups of models are enclosed within square boxes for clarity. The horizontal dotted line marks the zero level.}
    \label{fig:comparing Acoustic depths}
    
\end{figure}

Figure~\ref{fig:comparing Acoustic depths} shows the difference between $\tau_{\rm He}$, obtained from the glitch analysis, and the acoustic depth of the $\Gamma_1$ peak, $\tau_{\rm P}$, calculated from the outermost mesh point using the sound speed profile, for all models with $\chi_{2}^2 < 150$. Note the two groups of models: one with a smaller $\tau_{\rm He}$ of approximately 1300 s (bottom left box), and the other with values closer to the observed $\tau_{\rm He}$ listed in Table~\ref{tab:table_observed} for $\mu$ Herculis (top right box). The former group of models consistently appears in the upper left corner of the bottom panel of Figure~\ref{fig:glitch parameters}, exhibiting much higher values of $\Delta_{\rm He}$ (around 250 s) compared to the observed value. This trend is expected, as $\tau_{\rm He}$ and $\Delta_{\rm He}$ are anticorrelated, with a correlation coefficient of approximately -0.4, as indicated by the confidence ellipse. All models in this group have significantly higher chi-square values compared to several models in the other group, and they disappear from the diagram when we restrict the plot to models with $\chi_{2}^2 < 80$ (black dots). For these reasons, we do not consider these models representative of $\mu$ Herculis.

The group of models in Figure~\ref{fig:comparing Acoustic depths} with $\tau_{\rm He}$ values close to the observed value has $\Delta\tau = \tau_{\rm He} - \tau_{\rm P}$ of about 220 s. We attribute this discrepancy to the uncertainty in the calculation of $\tau_{\rm P}$. Note that $\tau_{\rm P}$ was calculated by performing the integral in Eq.~\ref{eq:acoustic_depth} from the $\Gamma_1$ peak to the outermost mesh point in the model. Like the photosphere, the outermost mesh point in a stellar model may also not correspond to the acoustic surface; in fact, for model M2, the outermost mesh point is just about 104 s above the photosphere. Thus, the actual acoustic depth of the peak for model M2 is larger than $\tau_{\rm P}$ by $394 - 104 = 290$ s, partly explaining the observed offset, $\Delta\tau \approx 220$ s. Furthermore, $\tau_{\rm He}$ should be compared with the location of the peak in the $\Delta\Gamma_1/\Gamma_1$ profile (see Eq.~\ref{eq:1}), which may differ from the location of the peak in the $\Gamma_1$ profile, particularly when the smooth background is not flat. To identify the smooth background, we fitted a third-degree polynomial to the regions surrounding the peak (see Figure~\ref{fig:gamma_1 profile}). For model M2, we identified the location of the peak in $\Delta\Gamma_1/\Gamma_1$, which lies approximately 51 s away from the $\Gamma_1$ peak in the direction of the photosphere. As we can see in Figure~\ref{fig:comparing Acoustic depths}, when we correct $\tau_{\rm P}$ for these, $\Delta\tau$ for model M2 decreases by $(394 - 104) - 51 = 239$ s and approaches zero. Table \ref{tab:table_sound} lists the acoustic depths, measured from the acoustic surface using the sound speed profiles, of the $\Delta\Gamma_1/\Gamma_1$ peak and dip for models M1 and M2, together with the depths obtained from the glitch analysis, $\tau_{\rm He}$, for comparison. Clearly, for these models, the values of $\tau_{\rm He}$ agree much more closely with the acoustic depth of the peak than with that of the dip. Although we did not compute the acoustic depth from the acoustic surface using the sound speed profile of the $\Delta\Gamma_1/\Gamma_1$ peak for all models due to the tedious nature of the process, tests with a few additional randomly selected representative models of $\mu$ Herculis indicate that $\tau_{\rm He}$ is consistent with the acoustic depth of the peak (and not with the dip). 

To further investigate whether the glitch analysis returns the properties of the $\Delta\Gamma_1/\Gamma_1$ peak or the dip, we compared the acoustic width obtained using the glitch analysis, $\Delta_{\rm He}$, with the acoustic widths of the peak and dip of models M1 and M2. As shown in the right panels of Figure~\ref{fig:gamma_1 profile}, the widths of the peak and dip were derived by fitting Gaussian functions to the corresponding $\Delta\Gamma_1/\Gamma_1$ profiles. They are listed in Table~\ref{tab:table_sound} along with $\Delta_{\rm He}$ for comparison. Clearly, for these models, the values of $\Delta_{\rm He}$ agree much more closely with the acoustic width of the peak than with that of the dip, consistent with the conclusion drawn in the preceding paragraph. Note that the model frequencies do not have statistical uncertainties, and hence the corresponding helium glitch parameters also do not have formal errorbars. Having said that, as mentioned earlier, the model frequencies were fitted using weights derived from the uncertainties on the observed frequencies for consistency. Assuming the same uncertainties on the fitted parameters in Table~\ref{tab:table_sound} as on the corresponding observed values, it is clear that the errorbars are sufficiently small (see Table~\ref{tab:table_observed}) to allow a clear distinction between the peak and the dip.

In the Appendix (see Section~\ref{sec:solar_model}), we repeat the above exercise for a solar model \citep[Model S;][]{chris96} and show that the values of $\tau_{\rm He}$ and $\Delta_{\rm He}$ inferred from the glitch analysis \citep{verma14b} are consistent with the acoustic depth and width of the peak in the $\Delta\Gamma_1/\Gamma_1$ profile, respectively.

It is noteworthy that the Gaussian fit to the $\Delta\Gamma_1/\Gamma_1$ peak is at least as good as that for the dip. Consequently, the analytical framework developed by \citet{houd07} remains applicable to the peak, yielding the same functional form as in Eq.~\ref{eq:2}. The distinction lies in the interpretation of the parameters: $A_{\rm He}$ now represents the area under the peak, $\Delta_{\rm He}$ and $\tau_{\rm He}$ correspond to its acoustic width and depth, respectively, and $\phi_{\rm He}$ acquires an additional phase shift of $\pi$ due to the positive nature of $\Delta\Gamma_1/\Gamma_1$ profile near the peak. 


\section{Conclusions}
\label{sec:conclusions}

We performed a detailed glitch analysis using the GlitchPy code and asteroseismic modelling employing the BASTA package of our target $\mu$ Herculis, a G5 subgiant star. We used eight seasons of the SONG-Tenerife radial velocity time series data (period 2014-2021) to measure its oscillation frequencies. A dense stellar model grid of 3000 evolutionary tracks in an appropriate parameter space, along with several other smaller grids, was calculated. We adopted two different fitting procedures: the first relied on fitting the spectroscopic constraints and the observed oscillation frequencies (Fit1); and the second was based on fitting the helium glitch constraints in addition to the aforementioned set of constraints (Fit2). 

We found a consistent mass and radius of $\mu$ Herculis from the two approaches, Fit1 and Fit2, which were also in good agreement with the G17 modelling results based on the initial release of seismic data. Interestingly, our age from Fit1, $8.8_{-0.4}^{+0.6}$ Gyr, was significantly higher than that found in G17, $7.8_{-0.4}^{+0.3}$ Gyr. Furthermore, the inferred helium abundance was (sub)solar despite the significantly supersolar metallicity observed for $\mu$ Herculis. 

When the helium glitch parameters were included as additional constraints (i.e., using the approach Fit2), the inferred age decreased to $8.4_{-0.1}^{+0.4}$ Gyr; however, it remained significantly higher than the value reported by G17. The difference in the inferred age could potentially arise broadly due to three factors: (1) differences in the observed frequencies, (2) variations in the input physics adopted in the stellar model grid calculations, and (3) differences in the specific modelling approaches employed. As expected for a super-solar metallicity star, $\mu$ Herculis, the inferred helium abundances increased to (super)solar values, $Y_s = 0.242_{-0.021}^{+0.006}$ and $Y_i = 0.290_{-0.028}^{+0.004}$, after the inclusion of the helium glitch constraints. For this reason, we recommend inferences based on Fit2.

We found a discrepancy exceeding 3$\sigma$ between the observed acoustic width and the corresponding model-predicted value. Additionally, a marginal difference of approximately 2$\sigma$ was observed between the model and observed values of the acoustic depth. To test whether these discrepancies arise due to our inappropriate choices of the input physics used in the calculations of stellar models, we computed several small grids with a different solar metallicity mixture (GS98), equation of state (FreeEOS), and including convective-core overshoot. The discrepancy between the observed and model acoustic widths persists, suggesting a potential shortcoming in the stellar evolution models. However, the use of the GS98 mixture (instead of the lower metallicity solar mixture AGS09) appears to improve the agreement between the observed and model acoustic depths, although a conclusive confirmation of this would require a more detailed analysis. 

We conducted a detailed investigation into the origin of the helium glitch by comparing the acoustic depths and widths of the $\Gamma_1$ peak, derived from model sound speed profiles, with those inferred from glitch analysis of the corresponding model frequencies for several representative models of $\mu$ Herculis. We found that the acoustic depth and width inferred from the glitch analysis correspond more closely to the location and extent of the peak in $\Delta\Gamma_1/\Gamma_1$ than to the dip. This result is further confirmed in Appendix~\ref{sec:solar_model} using a model representative of the Sun. Our results of the detailed analysis of the subgiant $\mu$ Herculis support the earlier findings of \citet{broom14} and \citet{verma14b} for red-giant and main-sequence stars, respectively. We conclude that the helium glitch analysis of a star’s observed oscillation frequencies yields properties of $\Gamma_1$ (more precisely $\Delta\Gamma_1/\Gamma_1$) peak between the two stages of helium ionisation. Like the dip, we showed that the peak also has a Gaussian profile, and hence the formalism of \citet{houd07} can be safely applied to it as well. Within this formalism, our results aid in an accurate interpretation of the helium ionization zone properties derived from the observed helium glitch signature. In particular, $A_{\rm He}$ quantifies the area under the $\Delta\Gamma_1/\Gamma_1$ peak profile, $\tau_{\rm He}$ specifies its location, and $\Delta_{\rm He}$ represents its width.


\section*{Acknowledgments}
KV acknowledges support from the Council of Science and Technology, Uttar Pradesh (project Id.: 3368 and sanction No.: CST/D-818). The support and the resources provided by PARAM Shivay Facility under the National Supercomputing Mission, Government of India at the Indian Institute of Technology, Varanasi are gratefully acknowledged. Funding for the Stellar Astrophysics Centre was provided by The Danish National Research Foundation (grant agreement no. DNRF106). We thank the referee, Dr. M. Farnir, for his careful reading of the manuscript and for providing insightful comments.

\section*{Data Availability}
The data underlying this article will be shared on reasonable request to the corresponding author.





\appendix


\section{Posterior probability distributions}

\begin{figure*}

    \centering
    \includegraphics[width=1.0\linewidth]{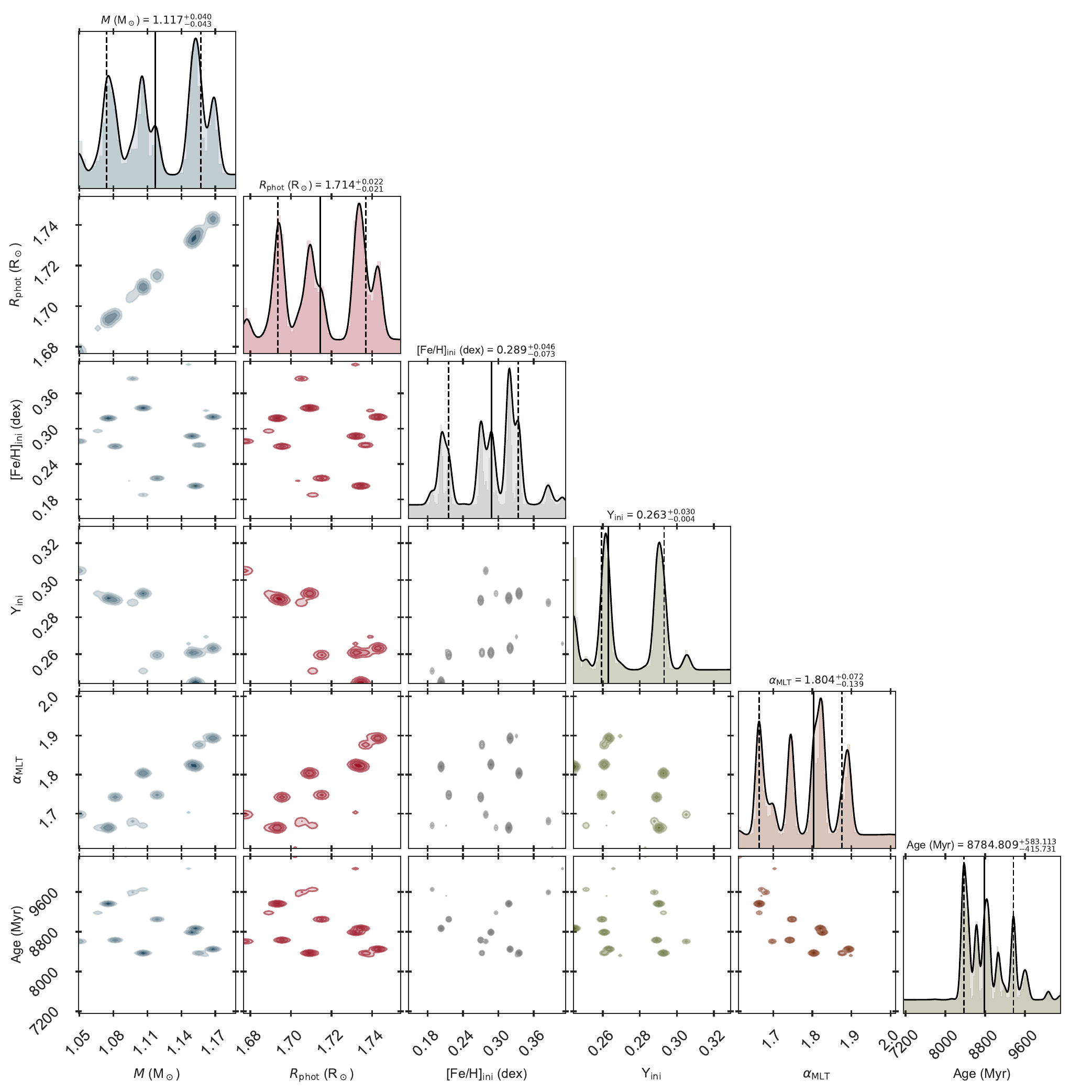}
    \caption{Corner diagram showing the posterior probability distribution obtained by fitting the observed effective temperature, surface metallicity, and oscillation frequencies (Fit1) of the $\mu$ Herculis.}
    \label{fig:corner_plot_freqs}
    
\end{figure*}

\begin{figure*} 

    \centering
    \includegraphics[width=1.0\linewidth]{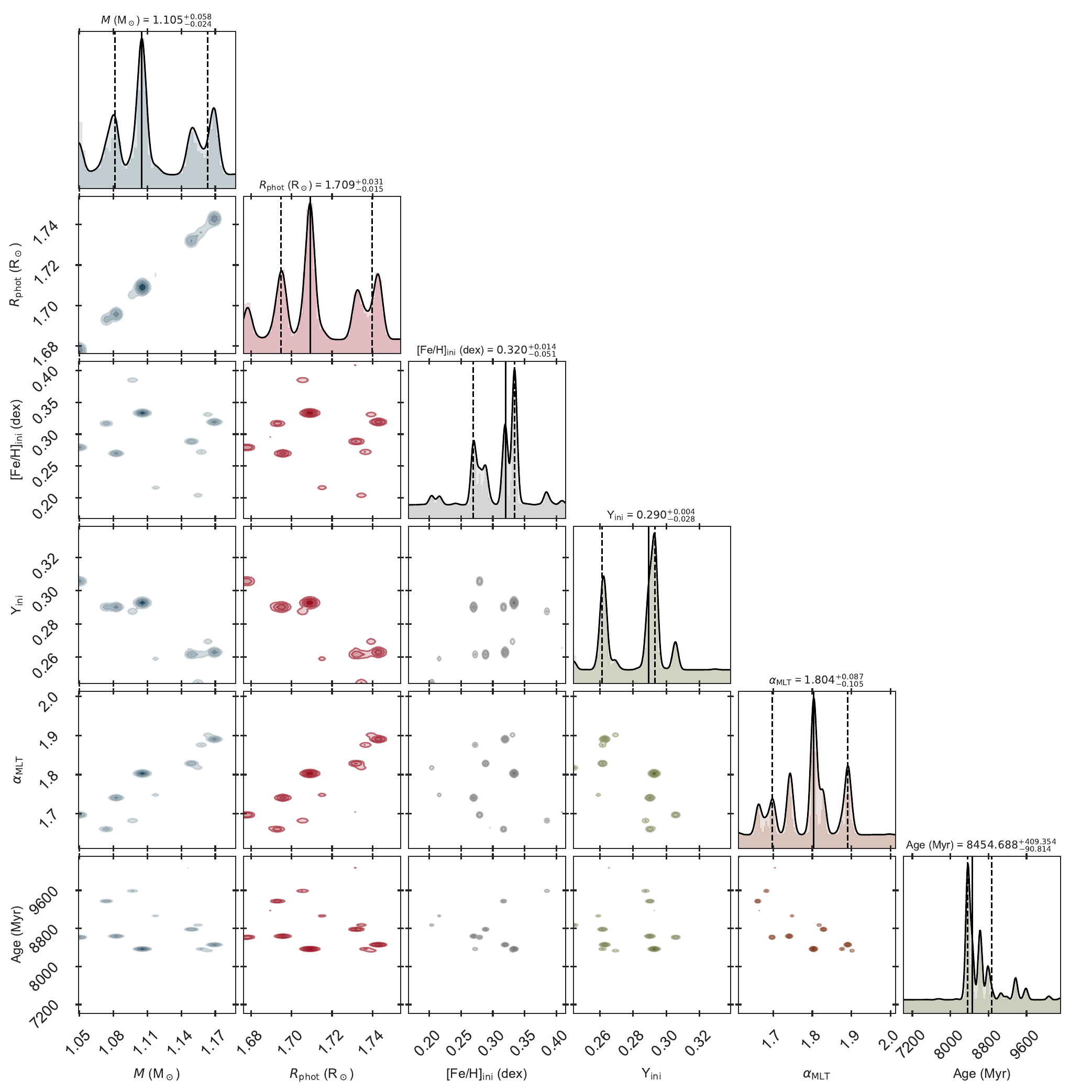}
    \caption{Corner diagram showing the posterior probability distribution obtained by fitting the observed effective temperature, surface metallicity, oscillation frequencies, and helium glitch parameters (Fit2) of the $\mu$ Herculis.}
    \label{fig:corner_plot_glitch}
    
\end{figure*}

We carried out two separate fits: (1) Fit1, which excludes the observed helium glitch properties; and (2) Fit2, which incorporates them. The posterior probability distributions resulting from Fit1 and Fit2 are illustrated in the corner diagrams shown in Figures~\ref{fig:corner_plot_freqs} and \ref{fig:corner_plot_glitch}, respectively. 

\section{Helium glitch in the Sun}
\label{sec:solar_model}

\begin{figure*}

    \centering
    \includegraphics[width=\linewidth]{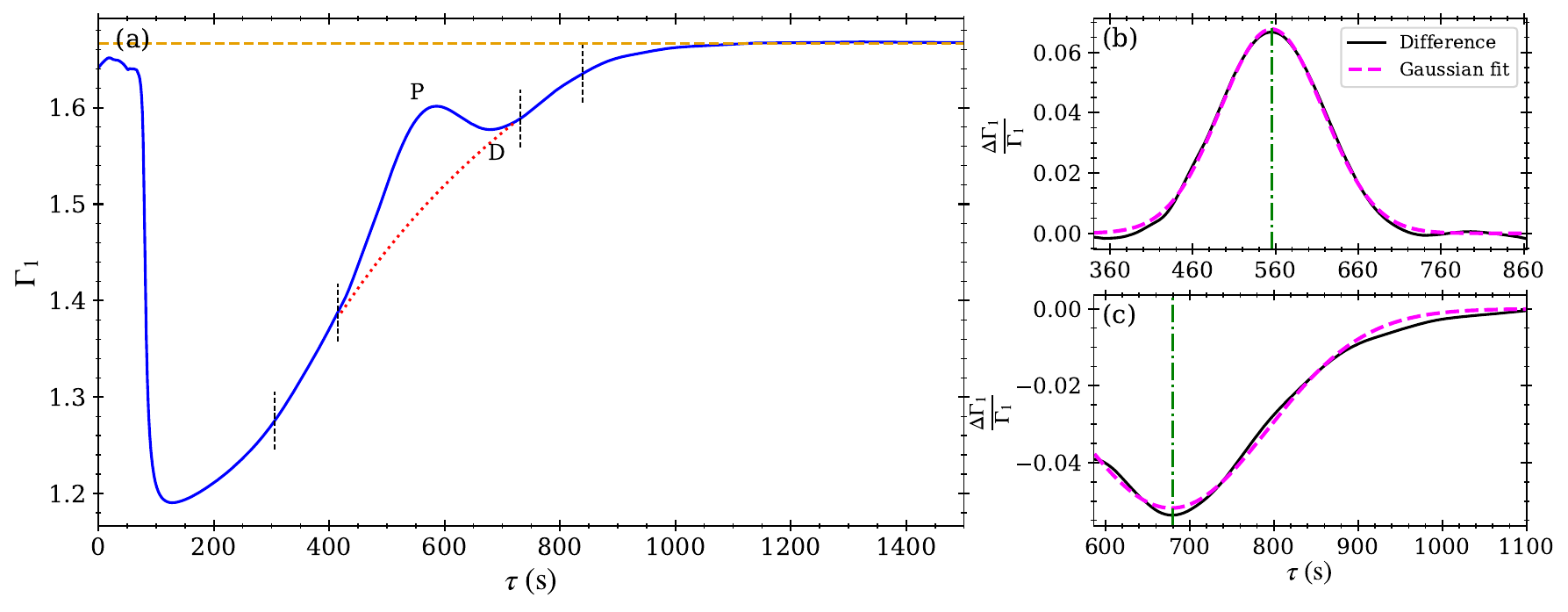}
    \caption{Same as Figure~\ref{fig:gamma_1 profile}, but for the solar Model S (see text for details).}
    \label{fig:gamma_1 profile_solar}
    
\end{figure*}

\begin{figure*}
    
    \centering
    \includegraphics[width=0.5\linewidth]{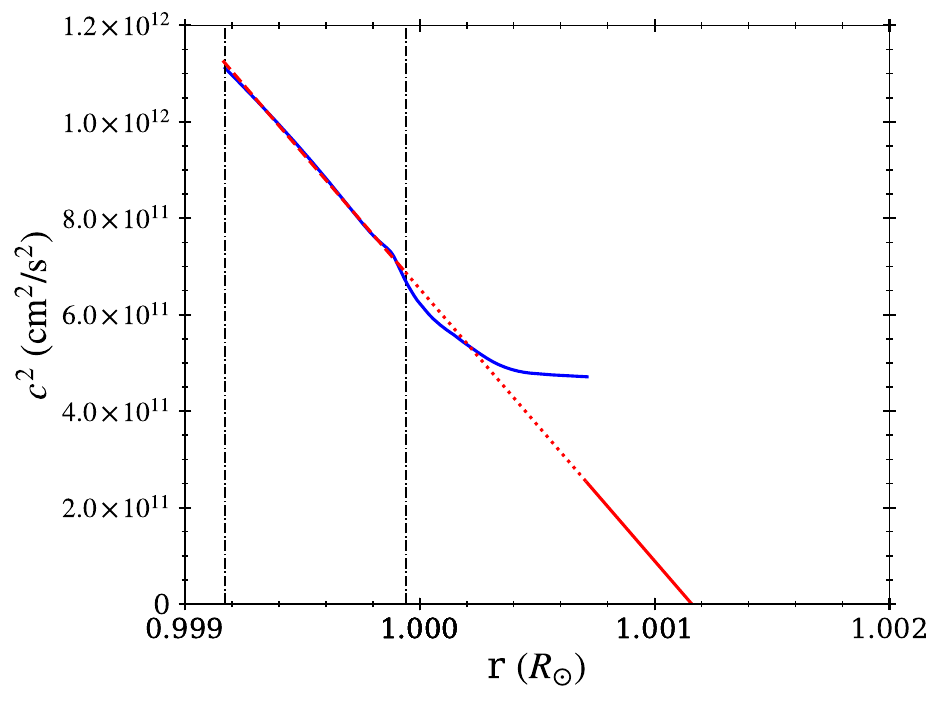}
    \caption{Same as Figure~\ref{fig:c2_muHer}, but for the solar Model S (see text for details).}
    \label{fig:c2_solar}
    
\end{figure*}

The observed acoustic depth and width of the helium ionisation zone for the Sun are listed in Table~\ref{tab:table_C1}. In the left panel of Figure~\ref{fig:gamma_1 profile_solar}, we show the $\Gamma_{1}$ profile of a solar model \citep[Model S;][]{chris96}. The acoustic depths of the photosphere, the $\Gamma_1$ peak, and the $\Gamma_1$ dip from the outermost mesh point are 69 s, 586 s, and 680 s, respectively. 

\begin{table}
    \caption{Same as Table~\ref{tab:table_sound}, except that it lists parameters for the Sun and a solar model. The column `Fitted’ lists observed values for the Sun from Table~1 of \citet{verma14b}.}
    \label{tab:table_C1}
    \centering
    \begin{tabularx}{\linewidth}{
        >{\centering\arraybackslash}p{2.5cm}|
        >{\centering\arraybackslash}X
        >{\centering\arraybackslash}X
        >{\centering\arraybackslash}X
    }
        \hline\hline
        \noalign{\vskip 0.1cm} 
        \raisebox{-0.25cm}{Parameters}\vspace{-0.2cm} & \multicolumn{3}{c}{Solar model}\\
        \noalign{\vskip 0.1cm}
        \cline{2-4}
        \noalign{\vskip 0.1cm}
        
        &Fitted& \hspace{-0.2cm} Actual (peak) & Actual (dip)\\
        \noalign{\vskip 0.1cm}
        \hline
        \noalign{\vskip 0.1cm}

        Acoustic depth~(s) & 696-707 & \hspace{-0.2cm} 678 & 802\\

        \noalign{\vskip 0.1cm}
        \hline
        \noalign{\vskip 0.1cm}
        
        Acoustic width~(s)& 60-61 & \hspace{-0.2cm} 62 & 114\\
        \noalign{\vskip 0.1cm}
        \hline
        \noalign{\vskip 0.1cm}
    \end{tabularx}
\end{table}`

We identified the acoustic surface in Model S (see Figure~\ref{fig:c2_solar}) and estimated the acoustic depth of the photosphere to be approximately 191 s. This value is slightly lower than 225 s reported by \citet{houd07}. Note that the value can vary by a few tens of seconds depending on the depth within the convection zone from which $c^2$ is linearly extrapolated. For the linear extrapolation of $c^2$, we identify the outermost convective layer and move inward by a distance equal to that between this layer and the outermost mesh point in the model. We identified the location of the peak in the $\Delta\Gamma_1/\Gamma_1$ profile, which lies approximately 30 s away from the peak in the direction of the photosphere. Since the smooth background is flat for the dip, the location of the dip remains the same in both $\Gamma_1$ and $\Delta\Gamma_1/\Gamma_1$ profiles. Therefore, the acoustic depth of the peak and the dip in the $\Delta\Gamma_1/\Gamma_1$ profiles from the acoustic surface are $586 + (191 - 69) - 30 = 678$ s and $680 + (191 - 69) = 802$ s, respectively (also listed in Table~\ref{tab:table_C1}). It is evident that the acoustic depth of the $\Delta\Gamma_1/\Gamma_1$ peak is in significantly better agreement with the observed value than that of the dip. 
 
We fitted Gaussian functions to the $\Delta\Gamma_1/\Gamma_1$ profiles, as shown in the right panels of Figure~\ref{fig:gamma_1 profile_solar}, to determine the widths of the peak and dip. They are listed in Table~\ref{tab:table_C1}. Once again, the width of the peak is consistent with the observed value, whereas that of the dip is not. 


\bsp	
\label{lastpage}
\end{document}